\documentclass{article}

\usepackage{arxiv}

\usepackage{cite}

\usepackage{mathtools}

\DeclarePairedDelimiter\ceil{\lceil}{\rceil}
\DeclarePairedDelimiter\floor{\lfloor}{\rfloor}
\usepackage{soul}
\usepackage{algorithm}
\usepackage{algpseudocode}

\usepackage{algorithm}
\usepackage{algpseudocode}
\usepackage{listings}

\usepackage{color}
\usepackage{xcolor}

{\normalfont\bfseries\tiny\color{red}}

\definecolor{delim}{RGB}{20,105,176}
\definecolor{numb}{RGB}{106, 109, 32}
\definecolor{string}{rgb}{0.64,0.08,0.08}

\lstdefinelanguage{json}{
    numbers=left,
    numberstyle=\small,
    frame=tb,
    captionpos=b,
    abovecaptionskip=5pt,
    belowcaptionskip=5pt,
    numbersep=-8pt,
    rulecolor=\color{black},
    showspaces=false,
    showtabs=false,
    breaklines=true,
    postbreak=\raisebox{0ex}[0ex][0ex]{\ensuremath{\color{gray}\hookrightarrow\space}},
    breakatwhitespace=true,
    basicstyle=\ttfamily\small,
    upquote=true,
    morestring=[b]",
    stringstyle=\color{string},
    literate=
     *{0}{{{\color{numb}0}}}{1}
      {1}{{{\color{numb}1}}}{1}
      {2}{{{\color{numb}2}}}{1}
      {3}{{{\color{numb}3}}}{1}
      {4}{{{\color{numb}4}}}{1}
      {5}{{{\color{numb}5}}}{1}
      {6}{{{\color{numb}6}}}{1}
      {7}{{{\color{numb}7}}}{1}
      {8}{{{\color{numb}8}}}{1}
      {9}{{{\color{numb}9}}}{1}
      {\{}{{{\color{delim}{\{}}}}{1}
      {\}}{{{\color{delim}{\}}}}}{1}
      {[}{{{\color{delim}{[}}}}{1}
      {]}{{{\color{delim}{]}}}}{1},
}

\title{Multi-GPU aggregation-based AMG preconditioner for iterative linear solvers}

\author{
  Massimo Bernaschi, Alessandro Celestini, Pasqua D'Ambra \\
  Institute for Applied Computing at National Research Council of Italy\\
  \texttt{\{name.surname\}@cnr.it} \\
   \And
  Flavio Vella \\
  University of Trento \\
  \texttt{\{flavio.vella\}@unitn.it} \\
}

\begin{document}

\maketitle


\begin{abstract}
We present and release in open source format a sparse linear solver which efficiently exploits heterogeneous parallel computers. 
The solver can be easily integrated into scientific applications that need to solve large and sparse linear systems on modern parallel computers made of hybrid nodes
hosting NVIDIA Graphics Processing Unit (GPU) accelerators. 

The work extends our previous efforts in the exploitation of a single GPU accelerator and proposes an implementation, based on the hybrid MPI-CUDA software environment, of a Krylov-type linear solver relying on an efficient Algebraic MultiGrid (AMG) preconditioner already available in the BootCMatchG library. 
Our design for the hybrid implementation has been driven by the best practices for minimizing data communication
overhead when multiple GPUs are employed, yet preserving the efficiency of the single GPU kernels. Strong and weak scalability results on well-known benchmark test cases of the new version of the library are discussed. Comparisons with the Nvidia AmgX solution show an improvement of up to 2.0x in the solve phase.
\end{abstract}

\section{Introduction}\label{sec:intro}
Heterogeneity is a feature of state-of-the-art computing platforms.
As a matter of fact, the trend toward using different types of processors in the same computing node is well established~\cite{S2019}.
From laptops to (pre-)exascale Top500 supercomputers~\cite{top500}, current computing platforms feature multi-core processors coupled with accelerators, like Graphics Processing Units (GPUs).
The efficient use of these heterogeneous systems, for general-purpose computation, requires the redesign of the algorithms and of their implementation, although some efficient and easy-to-use programming environments are able to hide most of the details of the architectures (e.g., Nvidia CUDA, SYCL or OpenACC just to cite few).
Exploiting all the computational power offered by such hybrid systems is particularly challenging for those applications that expose irregular memory access patterns, such as sparse linear solvers.

Let $A$ be a square, non-singular, matrix of (real) values and of dimension $n$, and $x$ and $b$ two column vectors of (real) values with consistent dimensions, a linear solver finds a solution to the problem $Ax=b$, where $A$ and $b$ are known and $x$ represents the unknown. An efficient and reliable linear solver, when matrix $A$ has many zero entries (often more than $99 \%$ of the overall) and very large $n$, is often the key element to obtain reliable simulations of complex models or to carry out big-data analysis. Most of the methods used to find solutions of sparse and large linear systems belong to the class of iterative Krylov-subspace projection methods, whose archetype is the well-known Conjugate Gradient (CG) method~\cite{Saadbook} applicable when $A$ is a symmetric and positive-definite (s.p.d) matrix. A basic component for the practicability of the CG method is the so-called preconditioner, that is a new (s.p.d.) matrix $B$ which approximates in some way the inverse of matrix $A$. The application of a preconditioner accelerates the convergence of the CG method when applied to the preconditioned system, such as $BAx=Bb$.

Some existing software tools, like Nvidia AmgX~\cite{AMGX2015,amgx} and hypre~\cite{hypre2016,hypre2021}, are able to exploit multi-GPU systems. In particular, AmgX includes Algebraic MultiGrid (AMG) preconditioners based on aggregation of unknowns driven by matchings in the adjacency graph of the system matrix, which are flexible methods able to balance the computational complexity and the convergence properties of the resulting linear solvers.
On the other hand, our AMG method implemented in {\em BootCMatchG}, leveraging on an aggregation scheme which exploits some properties of maximum weight matchings in graphs, offer better convergence properties in solving linear systems that fit in the memory of a single GPU~\cite{BDP2020102599,BDP2020100041}.
The main goal of the present paper is targeting at the scalability of that solution, making it able to solve sparse linear systems whose dimensions largely go beyond the memory capacity of a single computing node, such as those stemming from modern applications in Computational and Data Science.
To fill this gap, some approximations in the original numerical algorithms have been required, so that the combination of
communication-avoiding techniques, fine-grained parallelism, and overlapping between computation and data communication could allow us to design a scalable linear solver.
The contributions of this work are manifold:
\begin{itemize}
    \item a parallel AMG preconditioner for hybrid architectures able to show better convergence properties and scalability at lower complexity with respect to similar state-of-the-art AMG methods;
    \item a communication-efficient AMG preconditioner setup, based on parallel decoupled aggregation, which allows us to overlap computations of matching on GPUs with data communication needed for sparse matrix-sparse matrix products as well as to reduce overall data communication;
    \item a novel management of the data structure for non-local rows of sparse matrix in parallel hybrid sparse matrix-sparse matrix products which allows us to re-use at the best very efficient existing GPU kernels.
\end{itemize}
Our multi-GPU version of {\em BootCMatchG} is able to solve linear systems with more that $200$ millions of unknowns and scaling up to
$100$ hybrid nodes hosting Nvidia GPUs, showing a speedup with respect to AmgX, in the solve phase, ranging between $1.6$x and $2.0$x. 

The paper is organized as follows: Section~\ref{sec:HPC} briefly summarizes the main issues that affect the scalability and efficiency of scientific applications, and in particular of sparse linear solvers, on large-scale heterogeneous systems;
Section~\ref{sec:PCGsec} summarizes the features of~\emph{BootCMatchG} and of its components design, which are thought for efficient exploitation of highly parallel processors;
Section~\ref{sec:bootcmatchg} describes the main principles and guidelines that inspired our distributed algorithms on multi-GPU systems based on Nvidia GPU accelerators;
finally, Section~\ref{sec:results} presents strong and weak scalability results on realistic test cases stemming from the numerical solution of the 3D Poisson equation, which represents a ubiquitous kernel in many scientific applications. In particular, we compare our results with those obtained by a state-of-the-art solver, as implemented in Nvidia AmgX.

\section{Iterative Linear Solvers on heterogenerous HPC systems}
\label{sec:HPC}

Solving large and sparse linear systems is a ubiquitous and time-consuming task in Computational and Data Science applications. Physics-based simulations, statistical approaches for uncertainty quantification, and complex network analysis are only some of the application contexts. In particular, we focus on scenarios where the approximate solution is computed iteratively. Such an approach is generally used when the coefficient matrix $A$ is large and sparse or when the application tolerates a less accurate solution than that offered by the available machine precision.
In the context of heterogeneous computing, where we want to use
multi-core architectures and accelerators like GPUs to maximize performance, developing effective approaches for the iterative solution of linear systems brings the need of new methods, algorithms, and implementations capable of exploiting the underlying hardware and basic software. Indeed, regardless of the method adopted for solving the system, i.e., stationary iterative Jacobi and Gauss-Seidel or un-stationary Krylov methods~\cite{Saadbook}, the computation on sparse matrices is particularly challenging with respect to its dense counterpart due to the irregular memory access pattern and intrinsic load imbalance caused by the sparsity pattern of the matrix rows.

GPUs rely on fine-grained parallelism and access to medium-size memories with high bandwidth, but also not negligible latency. Therefore, although the current software stack makes available programming environments, e.g., Nvidia CUDA, which provide a clear interface to the features of the hardware, it is still challenging to use the above accelerators efficiently, especially for memory-bound kernels, like sparse matrix-vector products involved in iterative linear solvers.
For example, the Nvidia GPUs are built in terms of arrays of multithreaded, streaming multiprocessors, and each multiprocessor is composed of a fixed number of scalar processors. The CUDA programming paradigm is based on the concept of blocks of threads which share data. Therefore, having a regular density in the rows of a matrix is a favorable situation for high-throughput SIMD operations, whereas the irregular structure of general sparse matrices poses some limitations for efficient usage of the architecture which often are smoothed by organizing matrices in suitable data structures~\cite{FCBF2017}. Furthermore, a sequence of SIMD operations applied to the same data allows realizing the so-called data/thread-locality which makes GPU exploitation very efficient. For the same reasons, basic iterative algorithms which express a high-level of data parallelism are preferred to more complex algorithms which induce data dependency although, generally, the former may have worse convergence properties. Indeed, extra computation is often well tolerated and balanced by a very efficient execution on the GPU~\cite{DF2016,BDP2020102599}.
On the other hand, the ability to exploit scalability of large clusters of heterogeneous nodes largely depends on appropriate coordination among multiple levels of computation so that data partitioning, data/workload balancing, data communication between GPUs, CPUs, and among distributed nodes, do not penalize in a significant way the final performance.

During the years, many efforts have been devoted to design and implement efficient and reliable iterative sparse linear solvers in the form of open-source software libraries. Among the most widely used, we recall PETSc~\cite{petsc}, Trilinos~\cite{trilinos} and hypre~\cite{hypre2016,hypre2021}. All these software libraries went through many revisions to keep up with the technology development, as summarized in~\cite{Aetal2019}.
Sparse solvers relying on well-known AMG preconditioners specifically tuned for GPU clusters are available in Nvidia AmgX library~\cite{amgx}.

In this scenario, we propose an open-source software package for solving large and sparse linear systems with s.p.d. matrices as a fundamental component for an efficient exploitation of current high-end supercomputers.
Our preconditioning method, although requires a higher setup-time with respect to Nvidia Amgx solver, achieves a better scalability on multiple nodes and, generally, a faster convergence rate at a smaller complexity.
We combine the preconditioned CG (PCG) method, originally proposed in~\cite{NN2015} to be efficiently implemented on distributed-memory machines, and some of the enhancements introduced in~\cite{BDP2020102599,BDP2020100041} specifically designed for GPUs.

This work extends and improves our previous work by targeting very-large scale problems, having dimensions which go beyond the capacity of a single GPU. We apply an approach which introduces some approximations in the algorithms aimed at balancing high-levels of scalability with good convergence properties and computational complexity of the final sparse linear solver.

\section{Preconditioned Conjugate Gradient}
\label{sec:PCGsec}

Briefly, we recall the preconditioned CG (PCG) proposed by Notay~\cite{NN2015} that is particularly suitable for scaling on distributed machines.
It is a flexible version of the original CG, supporting the use of variable preconditioners, which relies on a re-organization of the basic operations, i.e., sequences of inner products and vector updates of the form $y=\alpha x + y$ (usually named {\em dot} and {\em axpy} operations, respectively), in order to reduce the overhead due to global synchronization and communication latency. The method is described in Algorithm~\ref{PCG}, where we highlight as the main computational kernels:
\begin{enumerate}
\item the application of the preconditioner represented by the matrix $B$ at lines~\ref{line:prec0} and~\ref{line:precB};
\item the sparse matrix-vector product ({\em SpMV}) involving the original matrix $A$ at lines~\ref{line:spmv0} and~\ref{line:spmv} ;
\item a sequence of {\em dot} products, instructions from line~\ref{line:stdot} to~\ref{line:enddot};
\item a sequence of {\em axpy} operations, instructions from line~\ref{line:staxpy} to~\ref{line:endaxpy}.
\end{enumerate}

The above version of the PCG brings out data locality, both in the sequence of {\em dot} products and in the sequence of the {\em axpy} operations, which can be also exploited at node-level for reducing global memory access, as we described in~\cite{BDP2020102599} and implemented in~\cite{BDP2020100041}.

%
%

\begin{algorithm}
\caption{Preconditioned Flexible Conjugate Gradient} \label{PCG}
\begin{algorithmic}[1]
\State Given $u_0$ and set $r_0 = b - Au_0$
\State $w_0 = d_0 = B r_0$ \label{line:prec0}
\State $v_0 = q_0 = A w_0$ \label{line:spmv0}
\State $\alpha_0 = w_0^T r_0$
\State $\beta_0 = \rho_0 = w_0^T v_0$
\Statex
\State $u_{1} = u_0 + \alpha_0 / \rho_0 d_0$
\State $r_{1} = r_0 - \alpha_0 / \rho_0 q_0$
\Statex
\For{$i = 1,\dots$}
	\State $w_{i} = B r_{i}$ \label{line:precB}
	\State $v_{i} = A w_{i}$ \label{line:spmv}
	\Statex
	\State $\alpha_i = w_{i}^T r_{i}$ \label{line:stdot}
	\State $\beta_i = w_{i}^T v_{i}$
	\State $\gamma_i = w_{i}^T q_{i-1}$ \label{line:enddot}
	\Statex
	\State $\rho_i = \beta_i - \gamma_i^2 / \rho_{i-1}$
	\Statex
	\State $d_{i} = w_{i} - \gamma_i / \rho_{i-1} d_{i-1}$ \label{line:staxpy}
	\State $u_{i+1} = u_i + \alpha_i / \rho_i d_{i}$
	\Statex
	\State $q_{i} = v_{i} - \gamma_i / \rho_{i-1} q_{i-1}$
	\State $r_{i+1} = r_i - \alpha_i / \rho_i q_{i}$ \label{line:endaxpy}
\EndFor
\end{algorithmic}
\end{algorithm}

\subsection{Algebraic Multigrid Preconditioner}
\label{sec:AMG}

One of the most relevant operations in the PCG is the application of a preconditioner which aims at reducing the number of iterations to solve the system within a given accuracy.
Specifically, a preconditioner is a matrix $B$ that attempts to improve the spectral properties of the system coefficient matrix. For s.p.d. matrices, it can be proved that the rate of convergence of the Conjugate Gradient method depends on the distribution of the eigenvalues of the system matrix~\cite{Saadbook}. It is expected that the condition number being smaller for the preconditioned matrix $BA$ w.r.t. the matrix $A$, i.e. 
$\kappa(BA)= \|BA\| \|(BA)^{-1} \| < \kappa(A)$, where $\| \cdot \|$ is a matrix norm.

From a computational perspective, there are two main issues that limit the scalability of preconditioning techniques.
The first one depends on the cost of the application of the preconditioner, that consists of multiplying the sparse matrix $B$ and the current residual vector $r_i$, as reported in Algorithm~\ref{PCG}, or in a more general procedure which computes the multiplication in an implicit way. The setup of $B$ is therefore particularly relevant; ideally its cost linearly increases in the problem size in terms of floating-point operations and memory footprint. It should be composed of as local as possible actions, reducing data communication among parallel tasks and leading to parallel performance which could be linearly increasing for an increasing number of tasks. We will refer to these features as {\em implementation scalability}.
The second aspect is related to the number of iterations that depends on the {\em optimal convergence} or {\em algorithmic scalability} of the preconditioner. An {\em ideal} preconditioner produces a preconditioned matrix whose condition number is very small ($\kappa(BA) \approx 1$) and independent of the problem size so that the PCG could have an almost constant number of iterations.

In this paper, we focus on a class of multigrid preconditioners named Algebraic MultiGrid (AMG) methods. They are considered purely algebraic since the information they use for building $B$ comes only from the system matrix $A$ rather than from the characteristics of the application domain. Such methods exhibit optimality both in computational complexity and convergence in some cases and therefore are suitable candidates for scalable parallel implementation~\cite{XZ2017,Vassilevskibook}.

The challenge in this type of preconditioners is the need to setup, in a purely algebraic setting, a hierarchy of coarse index spaces and matrices from the original system matrix (the {\em fine} matrix), which are sufficiently sparse and are still able to well represent, at each coarse level of the hierarchy, the components of the error vector which slow down the convergence of the PCG method.

In details, the setup of any AMG generates a hierarchy of $nl$ matrices, $A^k=(a^k_{ij})_{i,j=1, \ldots, n_k}, \ k=1, \ldots, nl$, where $A^1=A$ by a \emph{coarsening algorithm}.
For all $k < nl$, a prolongation matrix $P^k$ of dimension $n_k \times n_{k+1}$ is built and the matrix $A^{k+1}=(P^k)^TA^kP^k$ is computed according to the usual triple-matrix Galerkin product.

At each level of the hierarchy, a smoother $M^k$ is also defined, representing the iteration matrix of a simple relaxation method.
In our approach, we employ a weighted version of the Jacobi iterative method, named $\ell_1-$Jacobi. This method is highly parallel and always convergent for s.p.d. matrices, furthermore it exposes good smoothing properties for diagonally dominant matrices~\cite{BCHZ2013}.
Its iteration matrix is defined as the diagonal
matrix $M^k=diag(M_{11}^k, \ldots, M^k_{n_kn_k}),$ with $M^k_{ii}=a_{ii}^k+ \sum_{j \neq i} a_{ij}^k$, whose inversion is immediate.

The components produced in the setup may be combined in several ways to obtain different types of \emph{multigrid cycles} to be applied
in Algorithm~\ref{PCG} (line~\ref{line:precB})
as a procedure which computes the effect
of the application of the preconditioner operator to the residual vector.

An example of such a combination, known as V-cycle, is given in Algorithm~\ref{Vcycle_alg}.
In that case, a single iteration of the same smoother is used before and after the recursive call to the V-cycle (i.e., in the pre-smoothing and post-smoothing phases). At the coarsest level, i.e., for $k=nl$, a direct solver is usually employed. However, using direct solvers at the
coarsest level of an AMG method on multiple GPUs has two significant drawbacks: (i) the coarsest-level matrix tends to have a small size, therefore, in a multiple-node setting the cost of data communication dominates the local computations; (ii) direct solvers require the solution of a triangular system, which is particularly challenging on GPUs, due to the strong data dependencies and the load imbalance inherent to the sparse triangular structure~\cite{YRE2020}. In this work, we focus on a distributed iterative solver of the coarsest system to reduce data dependencies among parallel processors, in particular we apply the same version of the Jacobi method applied as smoother, for a fixed, sufficiently large, number of iterations. This approach allows us to re-use the kernels developed for the smoothing step and already demonstrated its benefits in the parallel version of the method for single GPU~\cite{BDP2020102599}.
\begin{algorithm}
\caption{\emph{V-cycle}} \label{Vcycle_alg}
\begin{algorithmic}[1]
\Procedure{V-cycle}{$k,nl,A^k,r^k$}
\If{ $k \ne nl$}
	\State $r^k = (M^k)^{-1} r^k $
	\State $r^{k+1} = (P^{k+1})^T r^k $
	\State $r^{k+1} =$ V-cycle$\left(k+1,nl,A^{k+1},r^{k+1}\right)$
	\State $r^k = P^{k+1} r^{k+1}$
	\State $r^k = (M^k)^{-T} r^k$
\Else
	\State $r^k = \left(A^k\right)^{-1} r^k$
\EndIf
	\State \textbf{return} $r^k$
\EndProcedure
\end{algorithmic}
\end{algorithm}
Algorithm~\ref{Vcycle_alg} corresponds to the application of a preconditioner matrix $B$ implicitly defined by the sequence of the following matrices:
\begin{multline*}
  B^k =  (M^k)^{-T}+(M^k)^{-1}-(M^k)^{-T}A^k(M^k)^{-1} +  \\
  (I-(M^k)^{-T}A^k)(P^kB^{k+1}(P^k)^T)(I-A^k(M^k)^{-1}) \ \ \  \forall k,
\end{multline*}
assuming that $B^{nl} \approx (A^{nl})^{-1}$ is an approximation of the inverse of the coarsest-level matrix.

Recently, we proposed an AMG relying on a coarsening algorithm based on a paiwise aggregation of unknowns which exploits maximum weight matchings (hereby {\em MWM}) in the adjacency graph of the system matrix. The graph weights are computed by a suitable function of a smooth vector (namely $w$) for the problem at hand and of the matrix entries. The procedure is applied recursively to obtain a hierarchy of coarser matrices until the coarsest matrix has a sufficiently small dimension ($coarsesize$) (see Algorithm~\ref{cmatch}).
\begin{algorithm}
\caption{\emph{Basic coarsening based on MWM}} \label{cmatch}
\begin{algorithmic}[1]
\Procedure{Coarsening}{$A,w,coarsesize$}
\State $k=1$, $A^1=A$, $w^1=w$
\While{ $size(A^k) > coarsesize$}
	\State apply {\em MWM}($A^k,w^k$)
	\State setup of pairwise prolongator $P^k$
	\State compute $R^k=(P^k)^T$
	\State compute $A^{k+1}=R^kA^kP^k$
	\State compute $w^{k+1}=R^kw^k$
    \State $k=k+1$
\EndWhile
    \State $NL=k$
	\State \textbf{return} $NL, A^2, \ldots, A^{NL}$
\EndProcedure
\end{algorithmic}
\end{algorithm}

The $NL-1$ paiwise prolongator operators built after matching correspond to piecewise constant interpolators having only one entry per row and two entries per column whose values depend on the chosen smooth vector.
More aggressive coarsening, with aggregates merging multiple pairs and having almost arbitrary large size of the type $2^s$ for a given $s>1$, can be obtained by combining multiple steps of the basic pairwise aggregation, i.e., by computing the product of $s$ consecutive pairwise prolongators; then, the final $nl-1$ prolongator operators have columns with at most $2^s$ entries. This choice of piecewise constant operators, also known as {\em un-smoothed} or {\em plain prolongators}, allows us to preserve sparsity of the coarser levels matrices, so that the final preconditioner has pretty small memory requirements and its application at each PCG iteration can be efficiently implemented in a parallel setting.
For details on the formulation and the mathematical rationale of our method we refer the reader to~\cite{DV2013,DFV2018,BDP2020102599,DV2019}. Here, we point out just that the main computational building blocks in the AMG preconditioner setup are:
\begin{enumerate}
\item the recursive application of {\em MWM} for building the sequence of basic pairwise prolongator operators $P^k$ for $k=1, \ldots, NL-1$;
\item triple-(sparse) matrix multiplications (hereby {\em SpMM}) to compute the coarse matrices $A^{k+1}=(P^k)^TA^{k}P^k$ at each level $k=1, \ldots, NL-1$;
    \item (sparse) matrix-(dense) vector products (hereby {\em SpMV}) to compute the coarse smooth vector $w^{k+1}=R^kw^k$ at each level $k=1, \ldots, NL-1$;
\item computation of a sequence of {\em SpMM} of the type $\overline{P}^1=P^1 \cdots P^s$, $\overline{P}^2=P^{s+1} \cdots P^{2s}$, $\ldots$, $\overline{P}^{nl-1}=P^{\floor{NL/s}s+1} \cdots P^{NL-1}$, to merge basic pairwise aggregates in aggregates of size at most $2^s$ and obtain the final hierarchy of $nl-1=\ceil{NL/s}-1$ prolongators;
\item computation of the new $nl-1$ coarse matrices by {\em SpMM} of Galerkin type involving the final sequence of prolongators $\overline{P}^k$.
\end{enumerate}
The last two steps may be not applied in the case of a final AMG hierarchy based on basic pairwise aggregates.

\section{BootCMatch on Multi-GPU systems}
\label{sec:bootcmatchg}

Enabling a hybrid MPI-CUDA scalable implementation of the computational procedure described in the previous section required a significant re-design of our single-GPU version. Our guidelines, as in the best practices, have been: efficient GPU exploitation by data and thread locality, minimization of data exchange between host and GPU accelerator as well as among multiple MPI tasks, possible exploitation of overlapping between computation and data communication among distributed MPI tasks.

The need to introduce parallelism in the operations for the setup of the AMG preconditioner led us to apply some algorithmic choices which impacted the quality of the coarsening algorithm described in Algorithm~\ref{cmatch}, as we detail in the following. However, as we demonstrate in Section~\ref{sec:results}, our approach is able to sufficiently preserve the convergence properties of the AMG preconditioner albeit some forms of approximation are introduced in its setup.
In the hybrid implementation, as in the original sequential code and in the mono-GPU version, the distributed sparse matrices are represented in the widely used Compressed Sparse Row (CSR) storage format. Furthermore, a row-block distribution of the matrices is applied, where each task owns a block of consecutive rows of the system matrix, and consecutive blocks of rows are assigned to MPI tasks with consecutive ranks.

\subsection{Decoupled Aggregation}
\label{subsec:graphmatch}

Our preconditioner setup
relies on a close-to-linear time complexity algorithm for computing matchings corresponding to a sub-optimal maximum weight. In general, algorithms for
computing matchings with maximum weight (optimal or exact matchings) are intrinsically sequential and feature a super-linear computational complexity~\cite{PFM2019}.  In~\cite{BDP2020102599} we discussed our effort in adapting an existing GPU implementation of the {\em Suitor} algorithm~\cite{NMHTL2015}, which computes a matching whose weight is, at least, half of the optimum. Although the {\em Suitor} aims at reducing the length of critical paths exploiting the principle of local dominant edges, it follows a bidding strategy to finalize matching. This bidding requires, in a distributed framework with multiple GPUs, a number of communication rounds, with corresponding expensive memory copies between CPU and GPU, that is basically equal to the number of tasks. It is apparent that, although it is possible to develop a multi-GPU version of the {\em Suitor} (and we did it for comparing
the quality of the results, see below), it can not be used in practice because it would hinder the scalability in a dramatic way. As a matter of fact, many research efforts are in progress to design parallel sub-optimal matching algorithms for very large graphs, as discussed in~\cite{Aceretal2021}. To the best of our knowledge, efficient algorithms and software exploiting multiple GPUs in a distributed setting are not yet available and their development is beyond the scope of this paper. However, for our aims, having a multi-GPU version of the MWM computation is not needed, as argued in the following.

A well studied parallelization strategy for coarsening in a distributed-memory programming model, which demonstrated to produce good quality coarse matrices in practice~\cite{TT2000,DDF2010,NN2015}, is based on a so-called {\em decoupled approach for the aggregation of unknowns}.
In details, every parallel task independently aggregates the subset of unknowns assigned to it in the row-block distribution of the current-level matrix. This decoupled aggregation is embarrassingly parallel, indeed it does not require any data
communication among the parallel tasks. However, it has a drawback which could affect the quality of the coarsening. Indeed it may produce non-uniform aggregates for the boundary unknowns, that are the unknowns owned by a
task and connected to unknowns owned by other tasks. Furthermore, it is dependent on the number of parallel tasks and on the initial partitioning of the system matrix.
We resort to the parallel decoupled approach of Algorithm~\ref{cmatch}, which results in the application of the {\em MWM} kernel only to the local subgraph corresponding to the local unknowns, ignoring the possible connections among unknowns owned by different tasks.
This choice allows us to exploit the already available mono-GPU version of the {\em Suitor} algorithm. Despite the above mentioned simplification, as also demonstrated by the results discussed in Section~\ref{sec:results}, the decoupled approach realizes a good trade-off between convergence of the final PCG and parallel efficiency of the preconditioner setup.
We observe that, after the decoupled {\em MWM} computation, since our aggregates involve only unknowns local to the GPU, the aggregation algorithm defines a distributed block prolongator (and then a restrictor) whose blocks, corresponding to the different tasks, can be locally built. As detailed in the next section, this special structure of the prolongation/restriction matrices also impacts on the successive computations to finalize the AMG preconditioner setup, i.e., on the parallel {\em SpMM} operations needed for possible products of successive prolongators and for the triple sparse matrix Galerkin products required to compute the coarser matrices.

\subsection{Parallel Hybrid SpMM Product}
\label{subsec:smsmprod}

General {\em SpMM} is a fundamental building block in several scientific computing applications involving sparse matrices and graphs, not only for AMG preconditioners in linear solvers. A large interest to have efficient {\em SpMM} kernels on parallel computers employing high-throughput accelerators has recently been observed in the context of Artificial Intelligence, e.g., in neural network compression using the factorization of layers into sparse matrix products~\cite{GBC2016,ZSGO2020,GAKAS2021}.

The main issue in {\em SpMM} is represented by the number of non-zero entries as well as the sparsity pattern of the resulting matrix product that are not predictable in advance.
A so-called {\em symbolic} phase is in general applied, in which the number of nonzeros in the result matrix is computed, postponing the actual calculations of the values to a following {\em numeric} phase. For the single GPU version of {\em BootCMatchG}, we resorted to the {\em nsparse}~\cite{NNM2017} package, which revealed much more efficient of any combination of primitives provided by Nvidia's {\em cuSparse} library. However, the {\em nsparse} solution had two disadvantages: $i)$ it relied on the legacy behaviour of CUDA {\em shuffle} primitives and $ii)$ it assumes that the two matrices are local to the GPU (there is no support for multi-GPU execution). While we solved the first issue by a simple updating of the shuffle primitives, according to their new definition introduced in CUDA-9, the second issue deserves a more detailed discussion.

Assumed that both operands are distributed in a consecutive row-block setting assigned to parallel tasks with consecutive MPI ranks, a straightforward solution to split the product computation among the GPUs is that each GPU computes the corresponding block of rows of the product matrix.
As an example, in the most simple configuration with just two GPUs, the first GPU is in charge of computing the first half of the rows and the second GPU is in charge of the second half of the rows of the product matrix. We describe the steps to be carried out for computing the rows of just the first half of the product matrix (i.e., from the viewpoint of the first GPU) but it will be apparent that the same approach can be applied to any subset of rows.
Each GPU owns the nonzeros of its row block of both the operands then, to finalize the scalar products of its rows of the first matrix by the columns of the second matrix, it generally needs the nonzeros of the second matrix corresponding to the row indices that are above or below its range of row indices, depending on the sparsity structure of each row of the first matrix. In our example the first GPU has just the nonzeros of the first half of each column of the second matrix. If a row belonging to its subset of the first matrix has nonzeros
whose column index is larger than $n/2$, where $n$ is the total number of rows (we assume that $n$ is an even number), it needs those nonzero entries to complete the product. In the following example of two very small ($4\times4$) matrices (the horizontal line represents the division between the two GPUs), the first GPU, in order to complete its part of the product, needs the elements, $b_{41}$ and $b_{44}$ of the last row in charge of the second GPU.
\begin{multline*}
\begin{bmatrix}
  a_{11}  & 0   & 0      & a_{14} \\
  0      & a_{22}   & 0      & a_{24} \\
  \hline
  0  & 0        & a_{33} & 0      \\
  a_{41}        & a_{42} & 0      & a_{44}
\end{bmatrix}
*
\begin{bmatrix}
 b_{11}  & 0        & 0      & b_{14}  \\
  0       & b_{22}   & b_{23} & 0 \\
  \hline
  0       & b_{32}   & b_{33} & 0      \\
  b_{41}       & 0   & 0      & b_{44} \\
\end{bmatrix}
= \\
\begin{bmatrix}
  a_{11} b_{11}+a_{14}b_{41}& 0 &0& a_{11} b_{14}+a_{14} b_{44} \\
  a_{24}  b_{41}& a_{22} b_{22} & a_{22} b_{23}&a_{24} b_{44} \\
  \hline \\
  \ldots \\
  \ldots
\end{bmatrix}
\end{multline*}

We exchange all the data necessary to complete the product on each GPU before starting the computation, so that the product appears as if it were completely local from the viewpoint of
the {\em nsparse} CUDA kernel. We made this choice because {\em nsparse} is very efficient up to the point that is better to call it once rather than to compute the local part of the product,
exchanging, in the mean time, the required data, then compute the remaining part of the product and sum the two contributions. This alternative choice could offer an advantage due to the
potential overlap of the computation of the local part with the exchange of the data for the non local part. However, it would entail a double execution of the {\em nsparse} kernel
(with two different symbolic steps) and also the execution of an additional kernel for the sum of the two partial products.
In summary, our final procedure for the multi-GPU {\em SpMM} is composed by the following steps, as sketched in Algorithm~\ref{alg:SpMM}:
\begin{enumerate}
\item Each GPU checks which rows of the second matrix needs to receive from other GPUs, (lines~\ref{startRTR} to~\ref{endRTR}).
\item CPUs exchange information about the number of non-zeros entries of each required row, using the following MPI collective communication primitives: {\tt MPI\_Allgather, MPI\_Alltoall, MPI\_Alltoallv}, (line~\ref{mpiCollective}).
\item CPUs, after the allocation of suitable memory buffers (whose size is set in the previous step), exchange the indices and the corresponding values of the non-local rows using MPI point-to-point {\tt (MPI\_p2p)} non blocking communication primitives, (lines~\ref{memAlloc} and~\ref{mpiP2P}).
\item Each GPU builds the subset of rows of the second matrix that it needs, by combining the rows it already owned with those received by other GPUs in the previous step. To minimize the number of memory copy operations, we introduced a new form of sparse matrix representation that we call {\em segmented} CSR. The idea is to maintain the local part of the matrix in its original CSR data structure and to store the non-local part, coming from other tasks, in an auxiliary CSR data structure.
The data structure of the auxiliary CSR should include, in general, all the information required to decide which
of the two parts, local or remote, is used during the computation of the matrix product. For the one-dimension decomposition adopted in the present work (i.e., each GPU has a block of consecutive rows), it is enough to know which subset of rows is local to the GPU. Each
column index (and corresponding value) within that range is accessed through the original CSR structure whereas any other column index (and corresponding value) is accessed through the auxiliary CSR. The original {\em nsparse} has been amended to include this simple criterion of selection whose cost is completely negligible with respect to the time saving it provides.
\item Each GPU carries out the product between its part of the first matrix and the suitable subset of the second matrix built in the previous step, (see line~\ref{rap} in Algorithm~\ref{alg:SpMM}).
\end{enumerate}
%
%
\begin{algorithm}
\caption{Parallel SpMM}\label{alg:SpMM}
\begin{algorithmic}[1]
\Statex \textbf{Input:} Two matrices $A$ and $B$, each process owns only a block of consecutive rows (from $h$ to $k$) of each matrix.
\Statex \textbf{Output:} Matrix $C = AB$, each process computes only
its block of consecutive rows (from $h$ to $k$).
\Statex
\State $rowsToReceive \leftarrow 0$
\For{$a_{ij} \in A_{local}$} \label{startRTR}
    \If{ $j < h$ \textbf{or} $j > k $}
        \State $rowsToReceive \leftarrow j$
    \EndIf
\EndFor \label{endRTR}
\State $nnzPerRow \leftarrow MPI_{coll}(nnz(B(rowsToReceive)))$ \label{mpiCollective}
\State Allocate memory for $B(rowsToReceive)$ combining the information contained in  $nnzPerRow$ and $rowsToReceive$ \label{memAlloc}
\State $B(rowsToReceive) \leftarrow MPI_{p2p}(B_{remote})$ \label{mpiP2P}
\State $B_{segmented} \leftarrow merge(B_{local}, B(rowsToReceive))$ \label{mSeg}
\State $C  \leftarrow sparseProduct(A_{local}, B_{segmented})$ \label{rap}
\State \textbf{return} $C$
\end{algorithmic}
\end{algorithm}
We observe that, in our implementation specifically tuned for the AMG preconditioner setup, steps 1), 4) and 5) are executed as GPU kernels, whereas steps 2) and 3) are carried out on CPU. This design choice is driven by the possibility to overlap computation and data communication, indeed the steps 2) and 3) can be executed on the CPU when the GPU is working on the {\em MWM} computation. As we discuss in section~\ref{subsec:weak}, this overlap allows us to completely hide the communication overhead when the number of parallel tasks is sufficiently small (see Fig.~\ref{fig-weak3}). However, when the number of tasks increases, the overlap is not longer balanced because the time
required by steps 4) and 5) becomes dominant with respect to the time required by the {\em MWM} computation (that remains always the same regardless of the number of tasks, see Section~\ref{subsec:graphmatch}). 

\begin{figure}[h!]
\begin{center}
\includegraphics[width=0.45\textwidth]{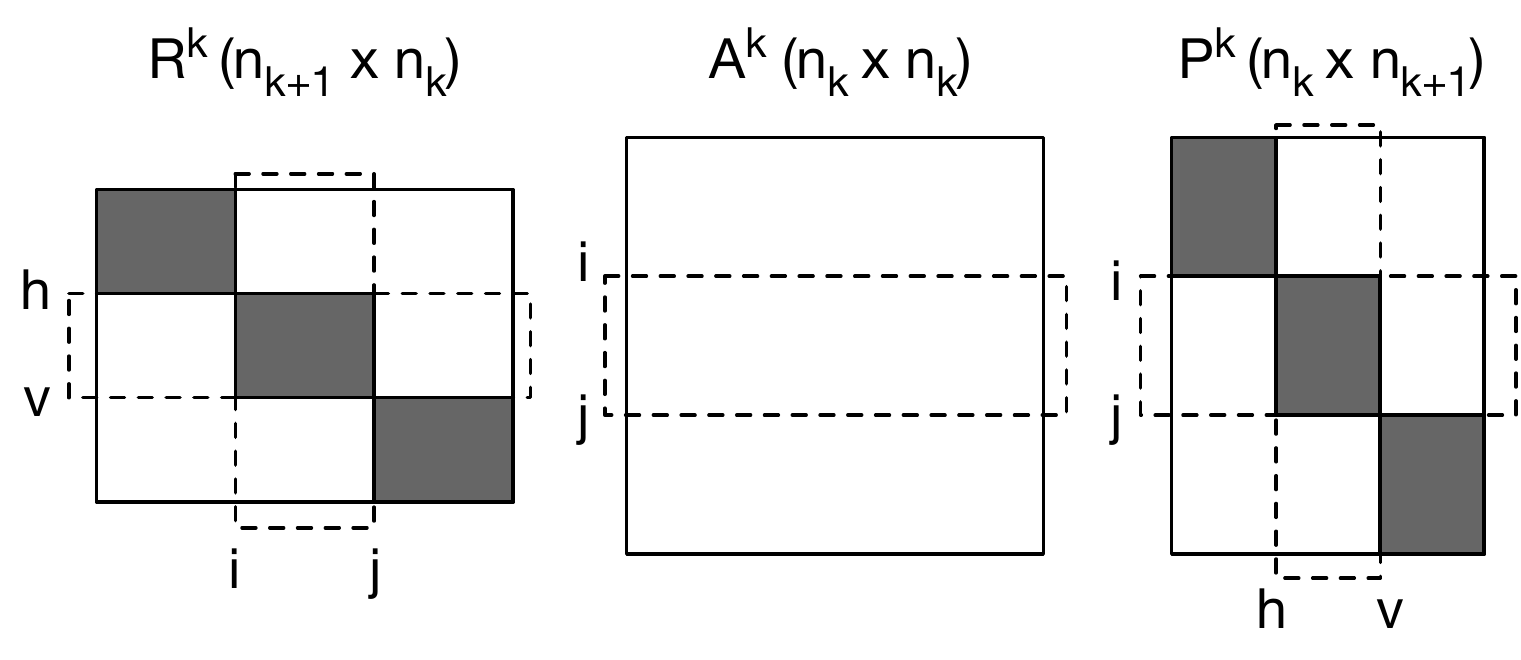}
\caption{Triple-matrix Galerkin product in the decoupled aggregation: matrices $R^k$ and $P^k$ have nonzeros only inside the gray blocks. \label{fig-rap}}
\end{center}
\end{figure}

We finally observe that, due to the decoupled aggregation approach described in Section~\ref{subsec:graphmatch}, we are able to completely avoid the communication for one {\em SpMM} operation of the triple-matrix Galerkin product needed to compute the coarse matrix at each level of the preconditioner hierarchy. Indeed, data communication among the tasks is only required for the first product $C=A^kP^k$, whereas the second {\em SpMM} product ($R^kC$) is completely local, because the rows of $R^k$ do not include nonzeros whose column indices correspond to rows owned by different tasks, as sketched in Figure~\ref{fig-rap}.

\subsection{Parallel hybrid SpMV Product}
\label{subsec:smvprod}

It is well known that sparse matrix-dense vector product is the key operation in iterative methods for linear solvers. In {\em BootCMatchG}, it is largely employed in the PCG method and in the V-cycle. As we described in~\cite{BDP2020102599}, depending on the sparsity pattern of the matrix, we resort to different solutions to carry out the product on the GPU. Regardless of the specific solution, when it comes to a distributed implementation using multiple GPUs, there is the need to exchange data among the GPUs to complete the product. The problem is similar to that described in Section~\ref{subsec:smsmprod}: any GPU may need some
entries of the vector owned by other GPUs. In the following example, the first GPU needs the entry $x_{4}$ whereas the second GPU needs the entries $x_{1}$ and $x_{2}$ to complete the corresponding {\em SpMV}.
\begin{equation*}
\begin{bmatrix}
  a_{11}  & 0   & 0      & a_{14} \\
  0      & a_{22}   & 0      & a_{24} \\
  \hline
  0  & 0        & a_{33} & 0      \\
  a_{41}        & a_{42} & 0      & a_{44} \\
\end{bmatrix}
*
\begin{bmatrix}
  x_{1} \\
  x_{2} \\
  \hline
  x_{3} \\
  x_{4} \\
\end{bmatrix}
=
\begin{bmatrix}
  a_{11} x_{1}+a_{14} x_{4} \\
  a_{22} x_{2}+a_{24} x_{4} \\
  \hline
  a_{33} x_{3} \\
  a_{41} x_{1} + a_{42} x_{2}+a_{44} x_{4} \\
\end{bmatrix}
\end{equation*}

It is apparent that the amount of data that needs to be exchanged among GPUs is very limited compared to the {\em SpMM} product. Each GPU needs to send, at most, $n/np$ entries to other GPUs (where $np$ is the number of parallel tasks) and needs to receive, at most, $\frac{np-1}{np}n$ entries from other GPUs. However, in the {\em SpMV} product, there is no need to carry out a symbolic step, so it makes sense to split the product in a local part that involves just the entries of the vector that each GPU already owns and a non-local part that involves the entries of the vector that each GPU needs to receive from other GPUs. This split makes possible to overlap the local computation with the communication of the necessary entries of the vector so that the communication overhead may be hidden. Albeit this approach is similar to what proposed in~\cite{KHWFBB2012,AMGX2015}, to the best of our knowledge, it has not been widely used in practice because its implementation is not straightforward. To provide an actual advantage, it requires several mechanisms: $i)$ asynchronous memory copies; $ii)$ CUDA streams; $iii)$ non-blocking MPI primitives that, properly combined, support the concurrent execution of memory copies, data-communication among CPUs and computation on local data carried out by the GPU. Our scheme is shown in Algorithm \ref{alg:SpMV}. 
The overlap is between the execution of the local part of the product (line \ref{LSPMV}) and the communication (the block between line \ref{MemcpyAsyncD2H} and line \ref{MemcpyAsyncH2D}). Then the execution is completed by computing the entries of the vector product that require communication with other tasks. With respect to \cite{KHWFBB2012}  we anticipate the execution of the local part exploiting the chance of running the kernel in line \ref{LSPMV} in asynchronous mode under the assumption that the output vector $y$ is different from the input vector $x$ (if the output vector had to replace the input vector, then a preliminary copy of the vector would be required).

\begin{algorithm}
\caption{Parallel SpMV\label{alg:SpMV}}
\begin{algorithmic}[1]
\Statex \textbf{Input:} A matrix $A$ and a vector $x$, each process owns only a block of consecutive rows (from $h$ to $k$) of matrix $A$ and the corresponding elements of vector $x$.
\Statex \textbf{Output:} Vector $y = Ax$, each process computes only
its set of consecutive elements of $x$ (from $h$ to $k$).
\Statex
\State $elToReceive \leftarrow 0$
\For{$a_{ij} \in A_{local}$} \label{startRTRMV}
    \If{ $j < h$ \textbf{or} $j > k $}
        \State $elToReceive \leftarrow j$
    \EndIf
\EndFor \label{endRTRMV}
\State $x(elToSend) \leftarrow MPI_{coll}(x(elToReceive))$ \label{mpiCollectiveMV}
\State Allocate memory for $x(elToReceive)$ and $x(elToSend)$ \label{memAllocMV}
\State $y  \leftarrow SparseMatrixVectorProduct(A_{local}, x_{local})$ \label{LSPMV}
\State Copy $x(elToSend)$ to CPU \label{MemcpyAsyncD2H}
\State $x(elToReceiveCPU) \leftarrow MPI_{p2p}(x_{elToSendCPU})$ \label{mpiP2PMV}
\State Copy $x(eToReceiveCPU)$ to GPU \label{MemcpyAsyncH2D}
\State $y  \leftarrow SparseMatrixVectorProduct(A_{local}, x_{remote})$ \label{RSPMV}
\State \textbf{return} $y$
\end{algorithmic}
\end{algorithm}
Within the iterations of the PCG method, the sparse matrices involved in the SpMV products do not change, therefore the indices which identify the vector entries to be exchanged in line \ref{mpiP2PMV} remain always the same. Finally, the pre-computation required for the communication operations (i.e., the selection of the elements to exchange among GPUs, corresponding to the block between lines \ref{startRTRMV} and \ref{memAllocMV} in Algorithm \ref{alg:SpMV}) is performed once before the solve phase and exploited throughout all iterations of PCG for each matrix in the AMG hierarchy.
We highlight that data communication is only needed for the {\em SpMV} products involving the original system matrix and the coarser matrices at each level of the AMG hierarchy, whereas all the {\em SpMV} products needed in the V-cycle and involving prolongator and restrictor operators are purely local because of the block structure of the related matrices.

\section{Experimental Setting and Results}
\label{sec:results}

In this section, we analyze parallel performance of the hybrid version of {\em BootCMatchG} on well-known benchmark test cases. We consider algebraic systems required for the solution of the Poisson equation in 3D:
\begin{equation}\label{eq:bench}
- \nabla \cdot ( K \nabla u ) = \mathbf{f}, \quad\text{ in } [0,1]^3,
\end{equation}
with homogeneous Dirichlet boundary conditions, $K= 1$ and unitary right-hand side.
The discretization of this problem is obtained by the classic 7-points finite-difference stencil for the left-hand side operator (the Laplacian operator), which results in the following system of algebraic equations:
\begin{multline*}
\frac{6 u_{i,j,k}}{h^2} \\ - \frac{u_{i-1,j,k} + u_{i+1,j,k} + u_{i,j-1,k} + u_{i,j+1,k} + u_{i,j,k-1} + u_{i,j,k+1}}{h^2} \\ = f_{i,j,k} \equiv 1,
\end{multline*}
for $i,j,k=1,\ldots,nd,$ and $h=1/(nd+1)^2$, and with $u_{i,j,k} = u(ih,jh,kh) \equiv 0$ on $\partial [0,1]^3$.
The above linear systems are characterized by a s.p.d matrix of coefficients and represent the computational kernel of many widely used scientific and engineering applications. Linear systems associated to the discrete Laplacian are also frequent in network analysis, e.g., in community detection algorithms based on spectral properties of the Laplacian graph associated with the network~\cite{DCV2019}.
We are interested in analyzing both {\em strong scalability}, i.e., the reduction in the execution times when a linear system with a fixed size is solved with an increasing number of parallel resources,
and {\em weak scalability} properties of our code for solving linear systems of increasing dimension, while parallel resources increase.
As already reported in Section ~\ref{sec:bootcmatchg}, data distribution is based on a row-block distribution of the system matrix among the MPI tasks and the related right-hand sides, that is blocks of contiguous rows are assigned to each task according to the order defined by the MPI rank.
We apply the hybrid version of \emph{BootCMatchG} to setup AMG hierarchies with the maximum size of the coarsest matrix fixed to $40 \cdot nd$. A maximum number of levels is also fixed to $40$. We compose couples of basic pairwise prolongator operators computed by basic pairwise aggregation, resulting in aggregates of size at most $8$, i.e., in prolongator operators having at most $8$ entries per column.
After completing the setup, we apply the AMG preconditioner as a single iteration of a V-cycle with $4$ iterations of $\ell^1-$Jacobi as pre- and post-smoother, whereas 20 iterations of the same smoother are applied at the coarsest level.
\par
We ran our experiments on Piz Daint\footnote{https://www.cscs.ch/computers/piz-daint/}, operated by
the Swiss National Supercomputing Center. It is based on the Cray XC40/XC50 architecture with $5704$ hybrid compute nodes (Intel Xeon E5-2690 v3 with
Nvidia Tesla P100 accelerator) and $1813$ multicore compute nodes (Intel Xeon E5-2695 v4), using the Cray Aries routing and communications ASIC with Dragonfly network topology. 
We ran the tests assigning $1$ MPI rank per each hybrid node embedding $1$ GPU.
\par
We analyze the performance of the linear solver looking at
the operator complexity $OPC$ of the resulting AMG hierarchy and $tsetup$, that is the execution time for the hierarchy setup. The number of iterations $it$ of the PCG, the execution time $tsolve$ to solve the system and the execution time $titer$ per each PCG iteration are also discussed to characterize the application phase from the viewpoint of both algorithmic and implementation scalability. All times are in seconds.
We observe that the operator complexity of an AMG hierarchy allows us to quantify both the memory requirements needed to store the operators at each level of the hierarchy and the cost of the application of a related V-cycle during the PCG iterations.
It can be measured by $OPC=\frac{\sum_{k=0}^{nl-1}\operatorname{nnz}(A^k)}{\operatorname{nnz}(A^0)} > 1$,
where $\operatorname{nnz}(\cdot)$ is the number of nonzero entries in a matrix.
Note that we stop PCG iterations when the relative residual in the $\ell_2$ norm is less than $10^{-6}$ or the number of iterations reaches the maximum value fixed to $1000$.

Our hybrid version of \emph{BootCMatchG}, hereby named {\em BCMG}, is compared with the hybrid version of Nvidia AmgX~\cite{amgx}. AmgX makes available various AMG preconditioners, based on different well known coarsening approaches already available in other libraries, such as hypre, and producing AMG hierarchies with different operator complexities. For a fair comparison, we selected the input configuration which defines AMG hierarchies based on aggregation of unknowns and having operator complexities comparable with the \emph{BCMG} preconditioner described above.
To this aim we modified few parameters of the configuration file named {\em PCG$\_$AGGREGATION$\_$JACOBI.json}, that is distributed with the library.
The selected AmgX preconditioner, hereby named {\em AMGX}, relies on a decoupled plain aggregation scheme based on local matching of unknowns as in our method, but their matching is driven by a simple heuristic. This heuristic, well understood only for the class of M-matrices~\cite{Vassilevskibook}, is a measure of strength of connections among the unknowns which derives from a characterization of the smooth error components; in analogy to what we do for \emph{BCMG}, aggregates of size $8$ are set. The resulting preconditioners are then applied in the same conditions described for \emph{BCMG}: at each iteration of the PCG method, $1$ iteration of a V-cycle, with $4$ pre/post-smoothing iterations of $\ell_1-$Jacobi at intermediate levels and $20$ iterations of $\ell_1-$Jacobi at the coarsest level, is applied. In all the cases, we built an AMG hierarchy with the same number of levels obtained by using our stopping criterion for coarsening in {\em BCMG}.
\subsection{Strong Scalability}
\label{subsec:strong}

\begin{figure*}[h!]
\begin{center}
\includegraphics[width=0.48\textwidth]{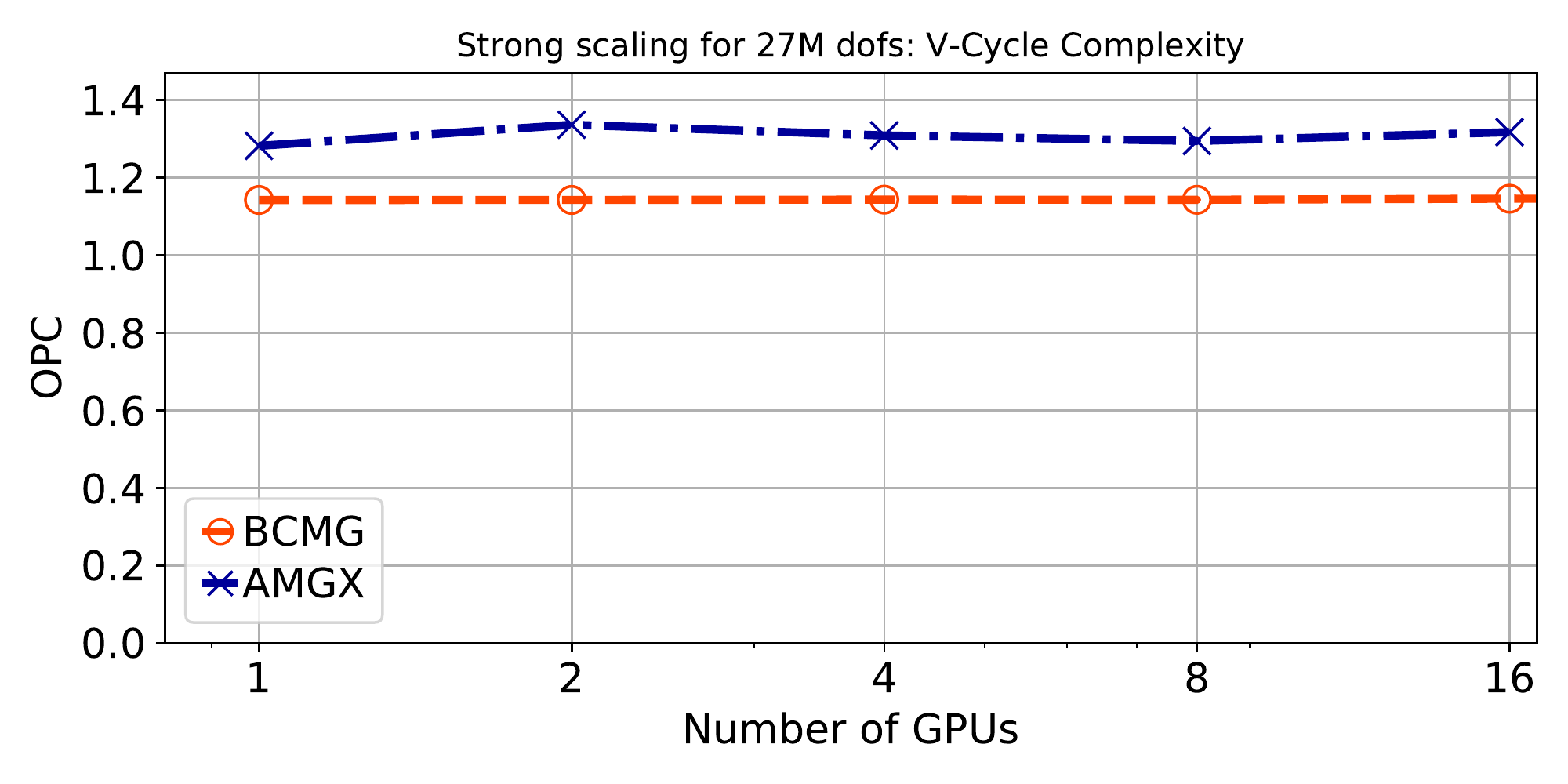}
\includegraphics[width=0.48\textwidth]{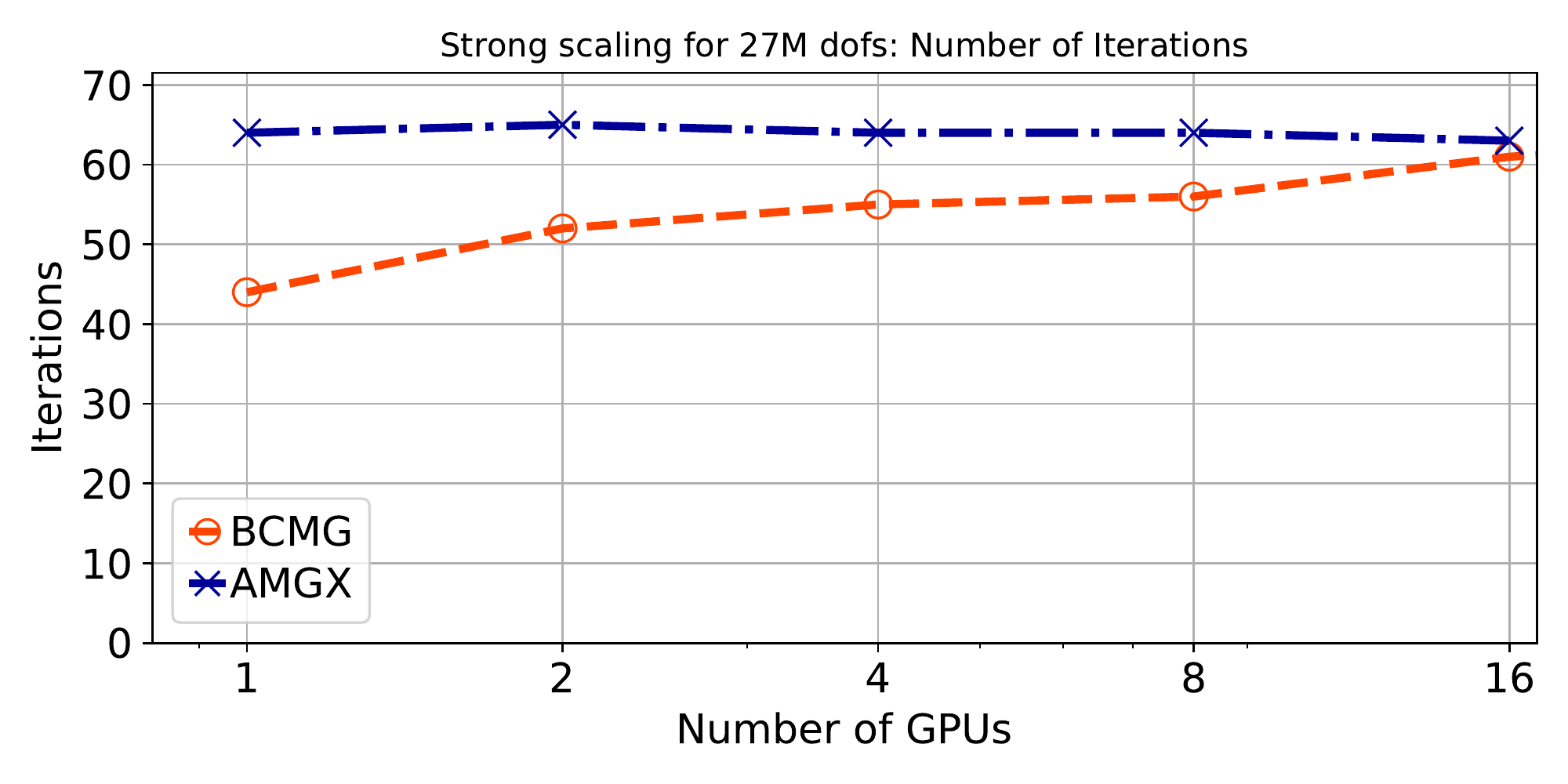}
\caption{Strong Scalability: Operator complexity and Number of PCG Iterations\label{fig-strong1}}
\end{center}
\end{figure*}

\begin{figure*}[h!]
\begin{center}
\includegraphics[width=0.48\textwidth]{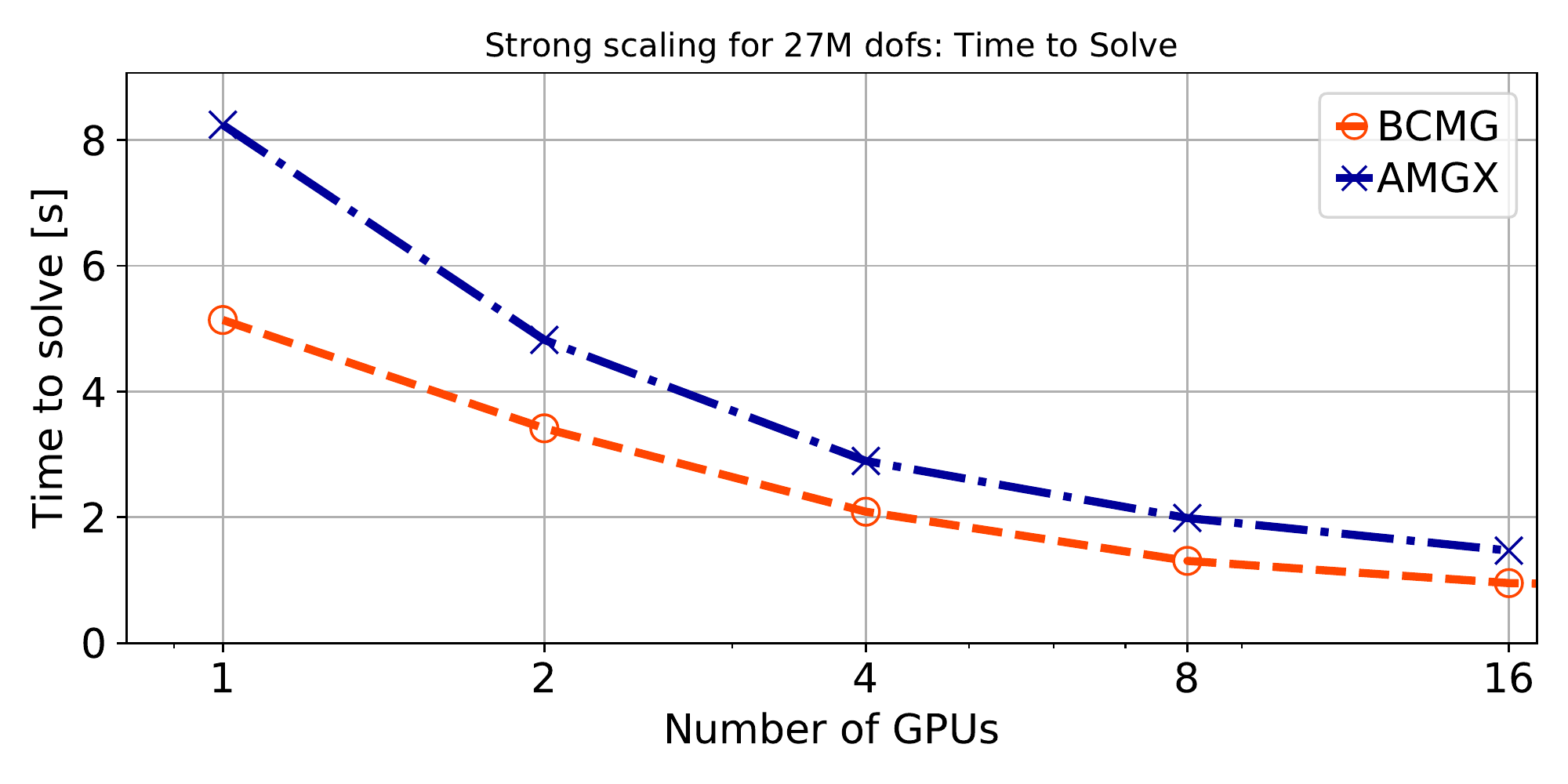}
\includegraphics[width=0.48\textwidth]{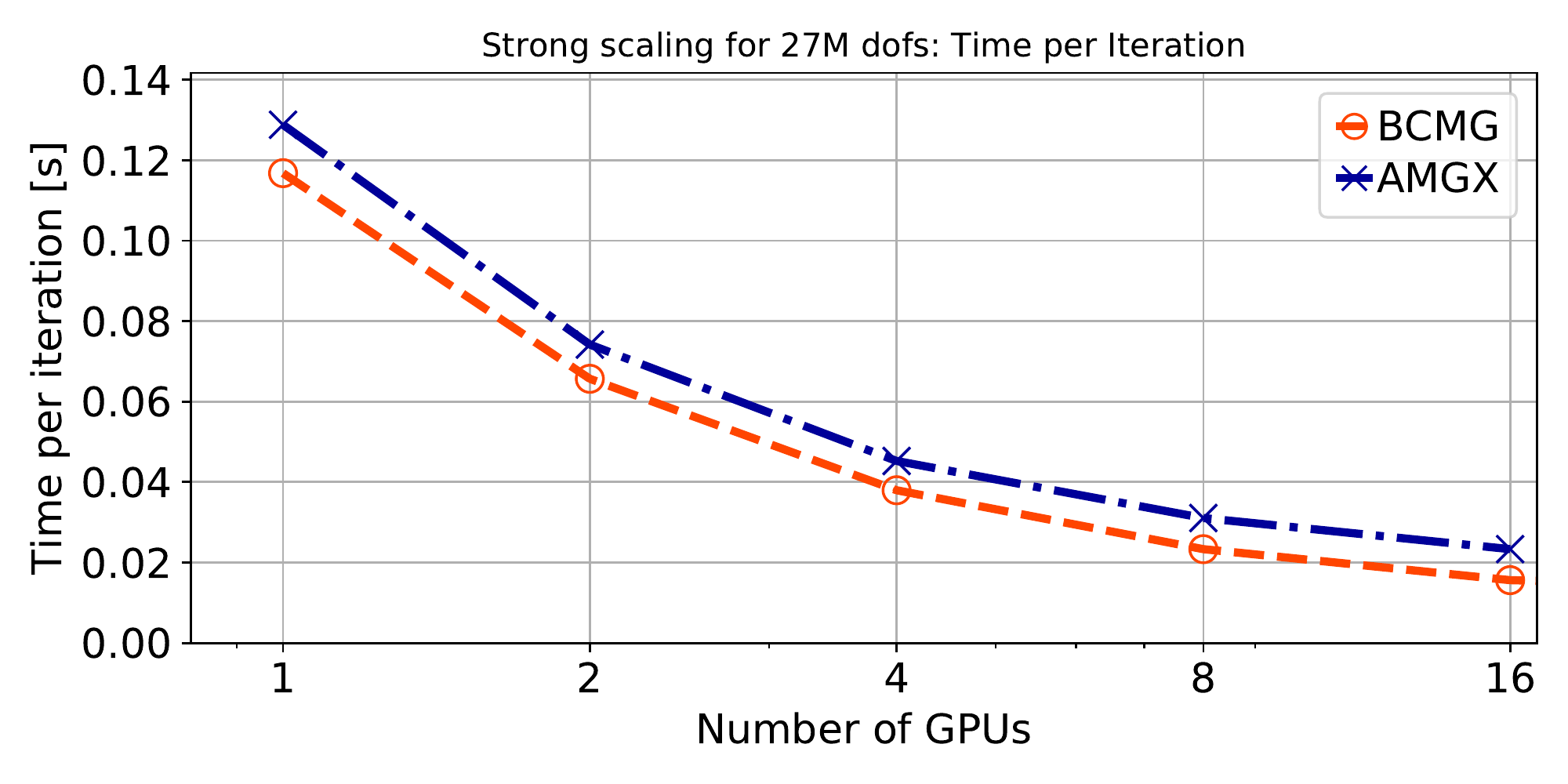}
\caption{Strong Scalability: Solve Time and Solve Time per Iteration\label{fig-strong2}}
\end{center}
\end{figure*}

\begin{figure*}[h!]
\begin{center}
\includegraphics[width=0.48\textwidth]{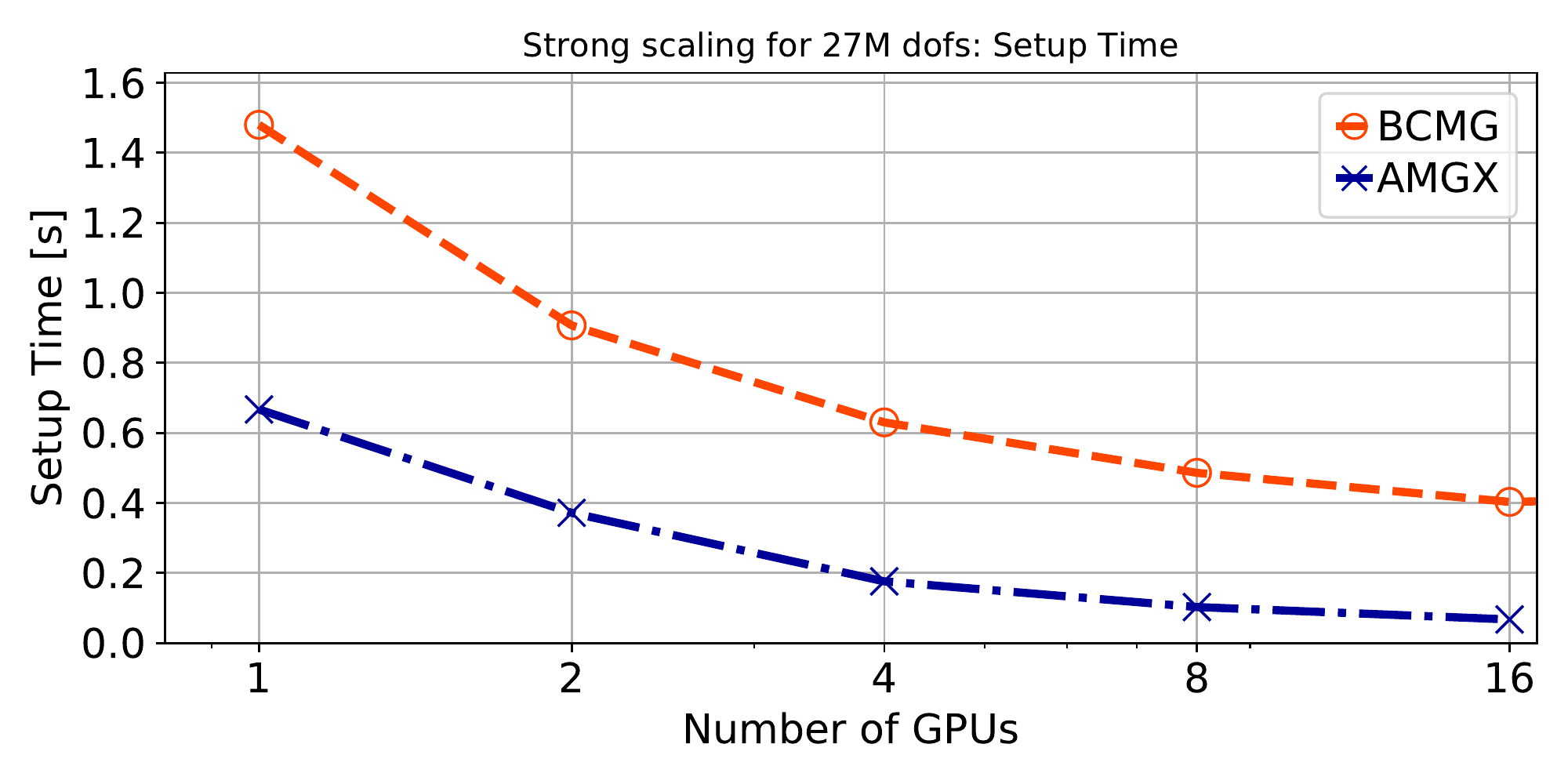}
\includegraphics[width=0.48\textwidth]{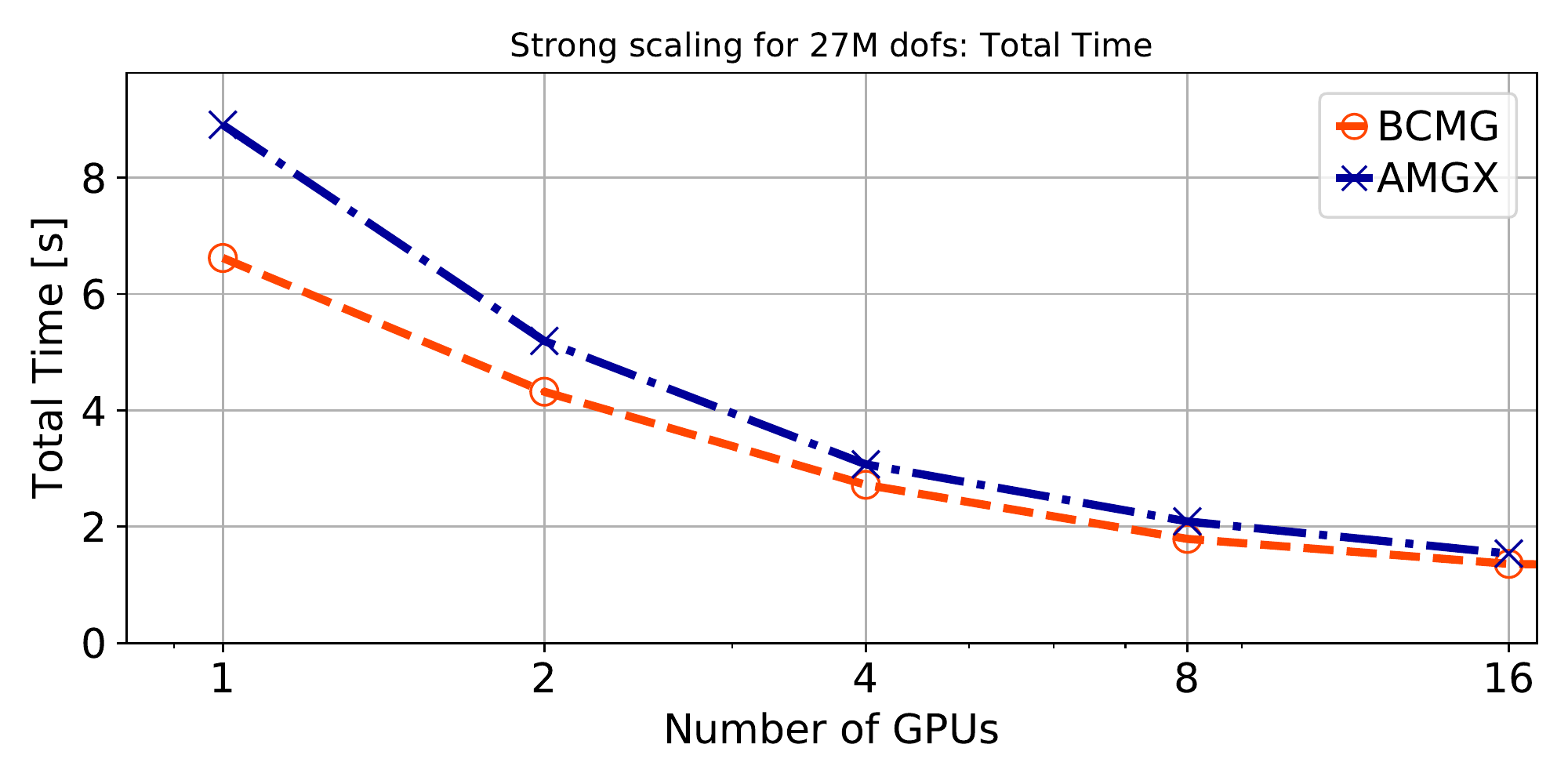}
\caption{Strong Scalability: Preconditioner Setup Time and Total Time\label{fig-strong3}}
\end{center}
\end{figure*}

We first discuss performance behaviour in solving a linear system with $300^3=27 \times 10^6$ degrees of freedom (dofs), increasing number of parallel nodes
embedding 1 GPU per node up to $16$ nodes. In all the cases, the number of AMG hierarchy levels is $5$.
In Fig.~\ref{fig-strong1}, we show operator complexity and number of iterations of {\em BCMG} and {\em AMGX}. We observe that in all the cases {\em BCMG} produces an AMG hierarchy with a smaller complexity than
{\em AMGX}; it is generally about $1.14$ for {\em BCMG}, whereas in the case of {\em AMGX} it ranges in the interval $[1.28, 1.34]$ corresponding to an increase in memory footprint and computational complexity of the preconditoner application ranging from $12 \%$ to $20 \%$ with respect to {\em BCMG}. On the other hand, we observe that in spite of the larger complexity, {\em AMGX} always requires a larger, albeit more stable, number of linear iterations in the solve phase. The expected variability in the number of iterations of {\em BCMG} for an increasing number of parallel nodes is due to a degradation of the quality of the coarse matrices related to the decoupled approach applied in Algorithm~\ref{cmatch}. Increasing
the number of parallel nodes corresponds to larger non-local matrix blocks which are ignored in the local MWM and in the corresponding block prolongators/restrictors; this is not a significant drawback because a large increase in the number of parallel resources for a fixed size problem is not beneficial due to the well known Amdahl's law effect.
The smaller operator complexity and number of iterations of {\em BCMG} corresponds to a smaller solve time going from $5.14$ sec. on $1$ node to $0.95$ on $16$ nodes (see Fig. \ref{fig-strong2}), which results in a satisfactory speedup of about $5.4$.
A slightly larger speedup of {\em AMGX} of about $5.60$ on $16$ nodes is essentially due to its very large solve time on $1$ node. This behaviour
highlights a very efficient implementation both at the node level and in exploitation of parallelism of all the computational kernels of the {\em BCMG} solve phase, as also confirmed by the very good behaviour of the time per iteration which shows a speedup of about $7.5$ on $16$ nodes, against a corresponding speedup of {\em AMGX} of about $5.53$.
For sake of completeness, in Fig. \ref{fig-strong3}, we also show the execution time for preconditioners setup and the total execution time given by the sum between solve time and preconditioner setup time.
As expected, the setup of {\em BCMG} requires a larger time due to the very different algorithmic approach. Indeed, our algorithm relies on the {\em MWM} kernel for coarsening and on an explicit triple sparse matrix product to generate the AMG hierarchy. On the contrary, the plain aggregation implemented in {\em AMGX} is driven by a simple heuristic based on a parameter-dependent measure of the connections' strength among the unknowns and builds binary prolongators so that the computation of the coarser matrices at each level requires just simple local sums. As a consequence, the overall procedure is embarrassingly parallel~\cite{AMGX2015}.
Nevertheless, a speedup of about $3.7$ is obtained for {\em BCMG} on $16$ parallel nodes, and the total execution time is always lower for {\em BCMG}.
\subsection{Weak Scalability}
\label{subsec:weak}

In this section, we analyze weak scalability of our solver; the fixed matrix size per node is equal to $130^3 \approx 2.2 \times 10^6$, going from $1$ to $100$ nodes. Therefore, we solve problems up to $ \approx 2.2 \times 10^8$ dofs. We observe that, in this case, the number of levels of the AMG hierarchies increases for increasing number of parallel nodes, ranging from $4$ levels on $1$ node to $5$ levels from $2$ to $16$ nodes, and finally to $6$ levels from $32$ up to $100$ nodes. This is due to our stopping criterion for the coarsening algorithm which tries to realize a good trade-off between the number of hierarchy levels and the size of the coarsest matrix, so that we can obtain a sufficiently accurate solution of the coarsest system at each preconditioner application while preserving parallel efficiency.
\par
\begin{figure*}[h!]
\begin{center}
\includegraphics[width=0.48\textwidth]{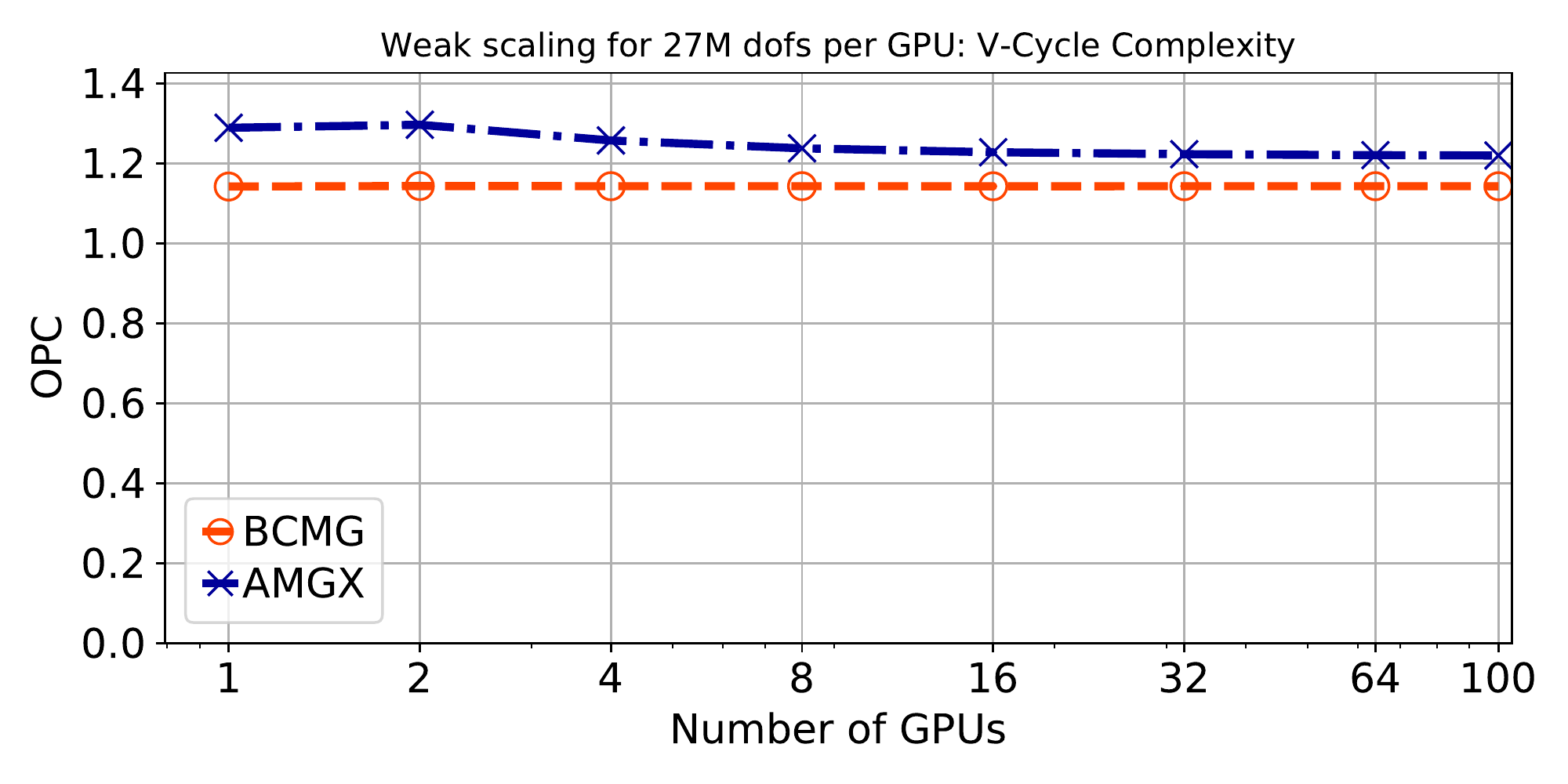}
\includegraphics[width=0.48\textwidth]{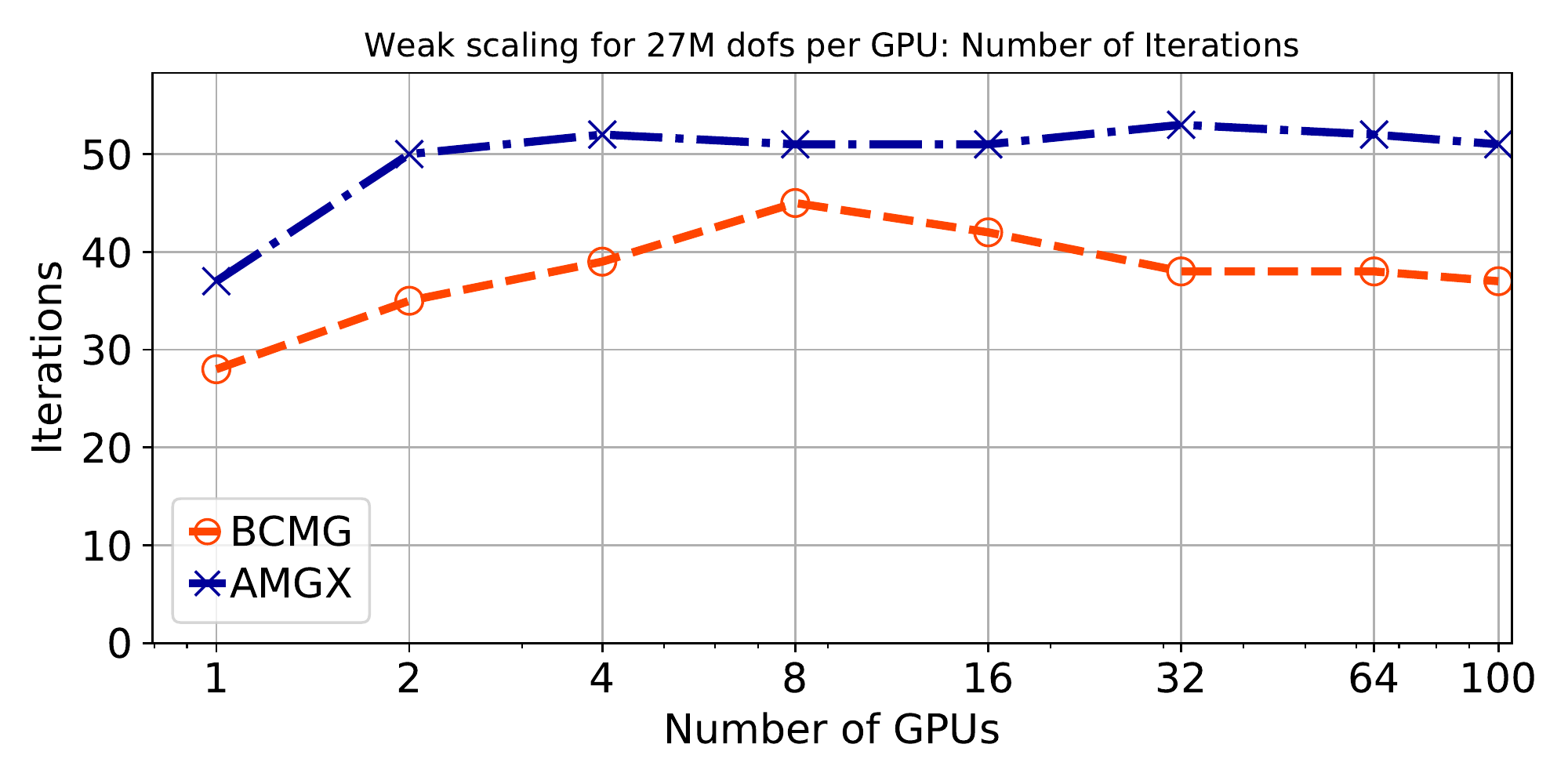}
\caption{Weak Scalability: Operator complexity and Number of PCG Iterations\label{fig-weak1}}
\end{center}
\end{figure*}
\begin{figure}[h!]
\begin{center}
\includegraphics[width=0.48\textwidth]{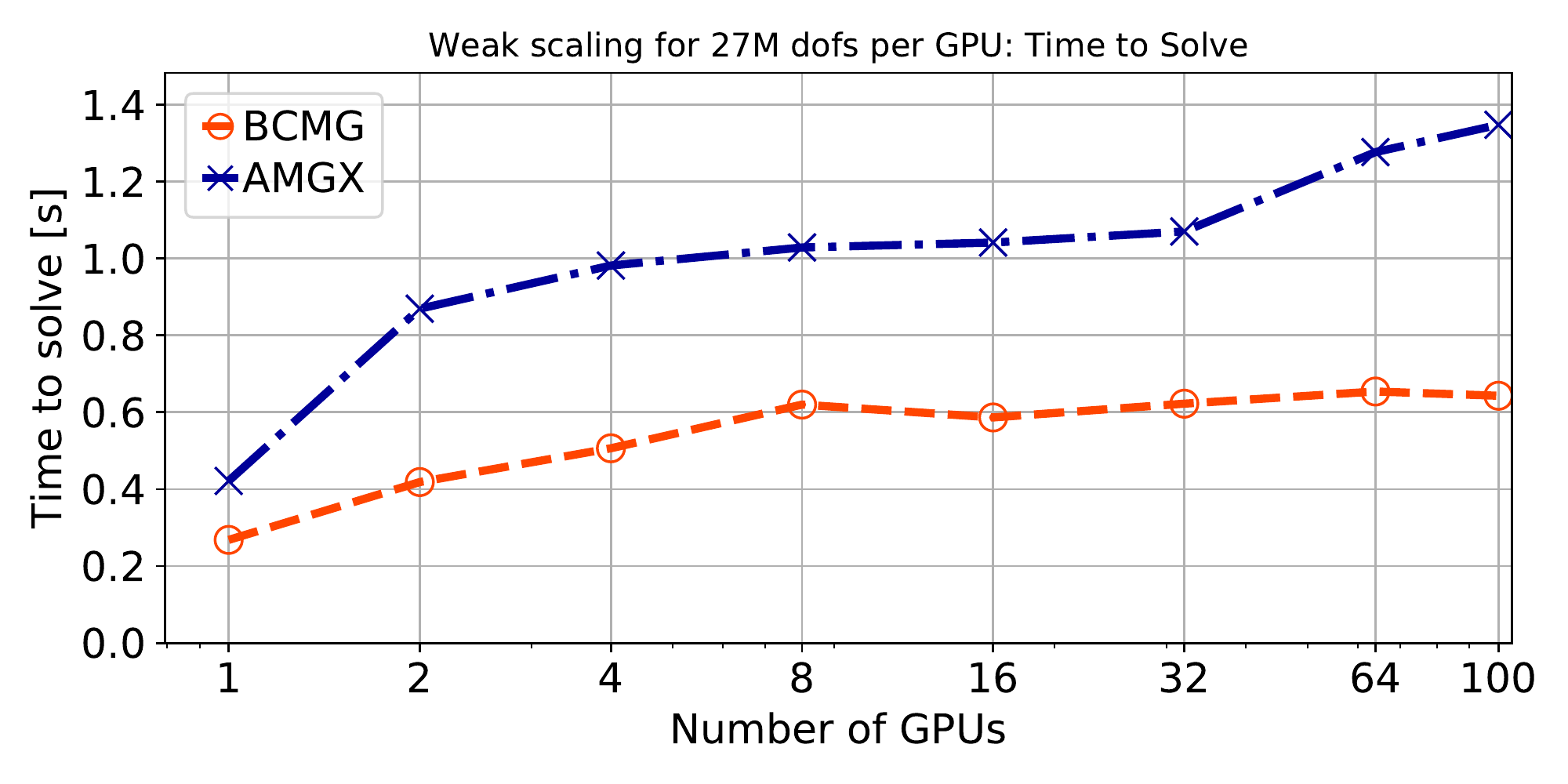}
\includegraphics[width=0.48\textwidth]{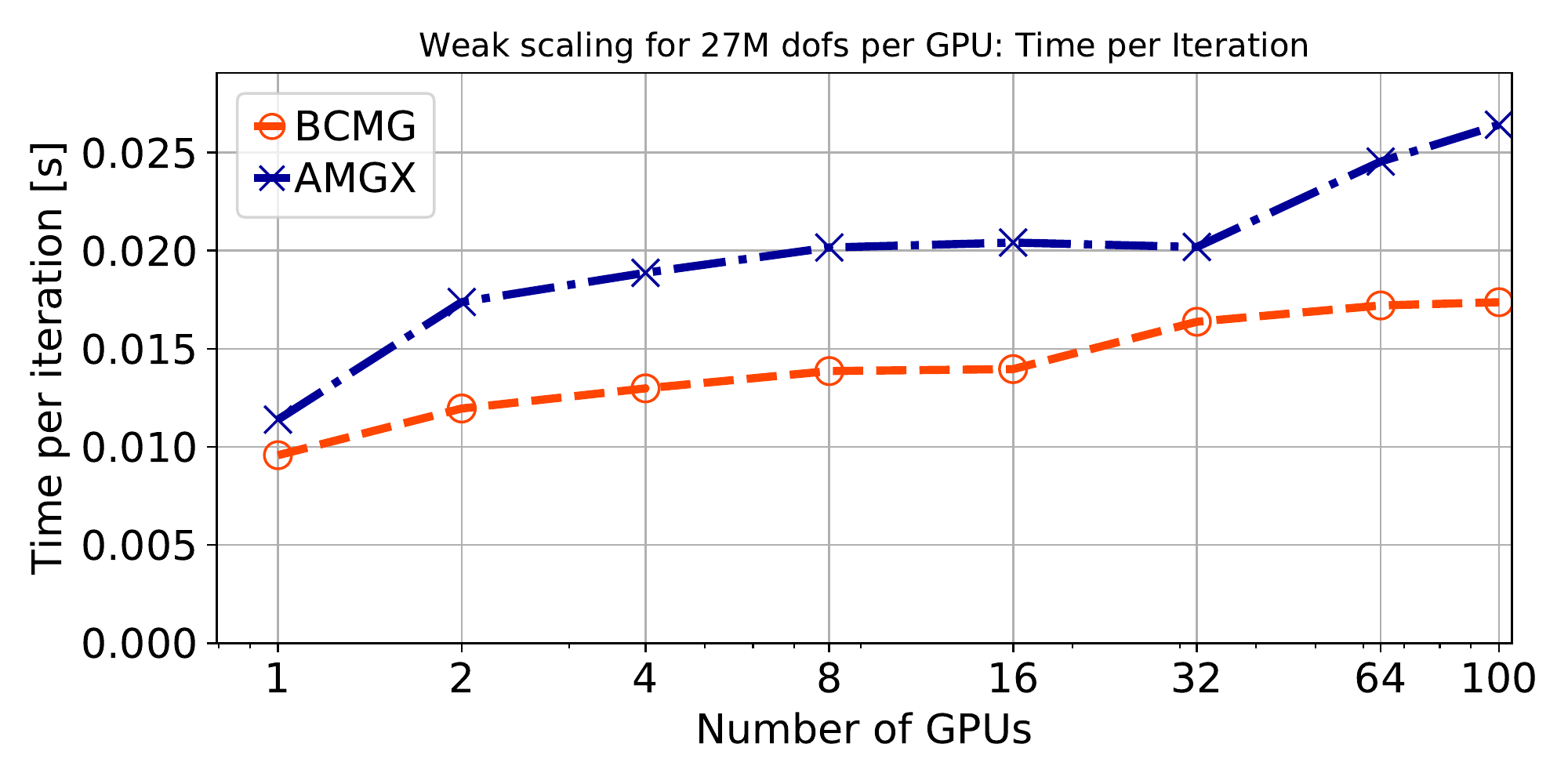}
\caption{Weak Scalability: Solve Time and Solve Time per Iteration\label{fig-weak2}}
\end{center}
\end{figure}
\begin{figure*}[h!]
\begin{center}
\includegraphics[width=0.48\textwidth]{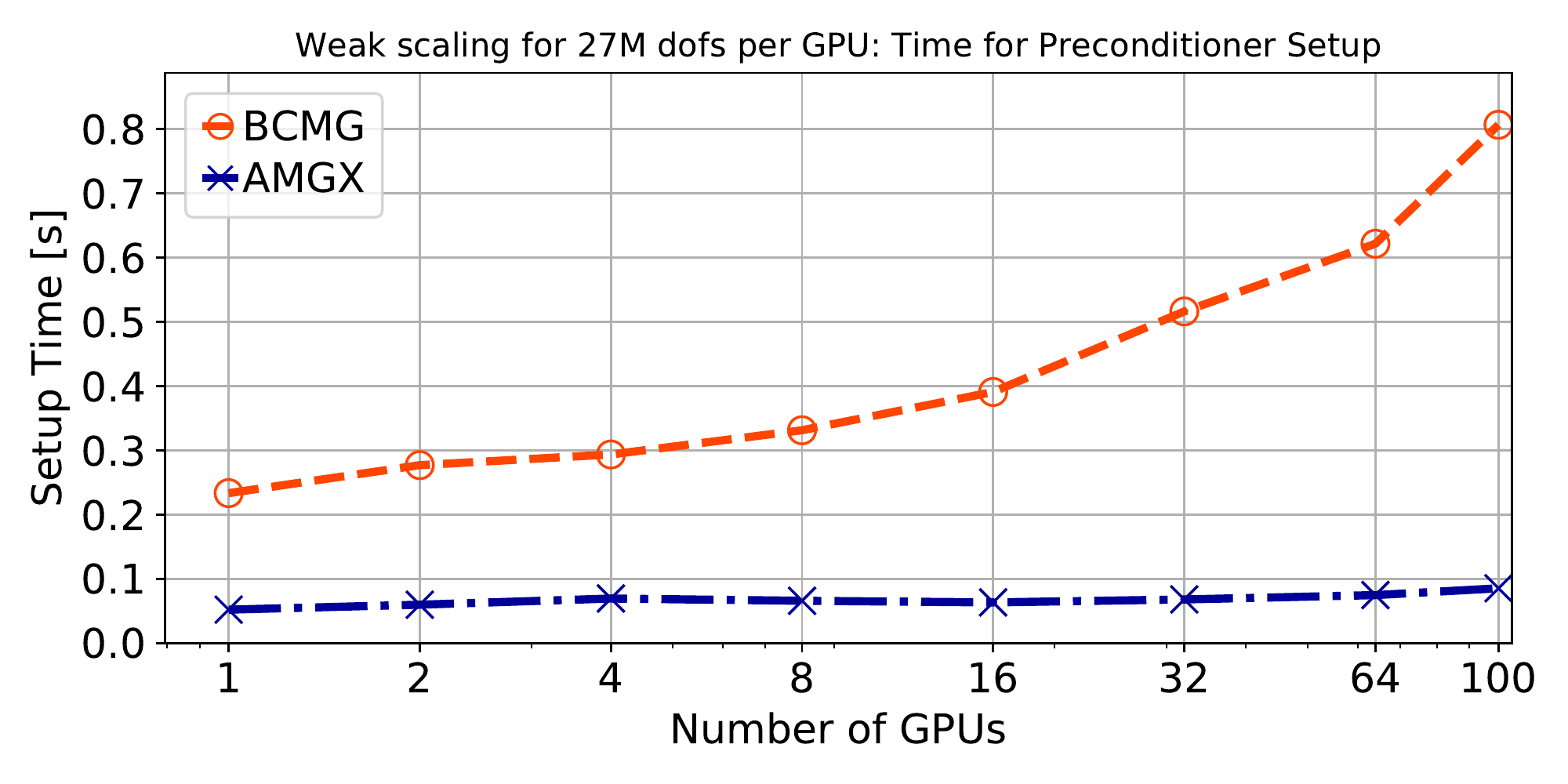}
\includegraphics[width=0.48\textwidth]{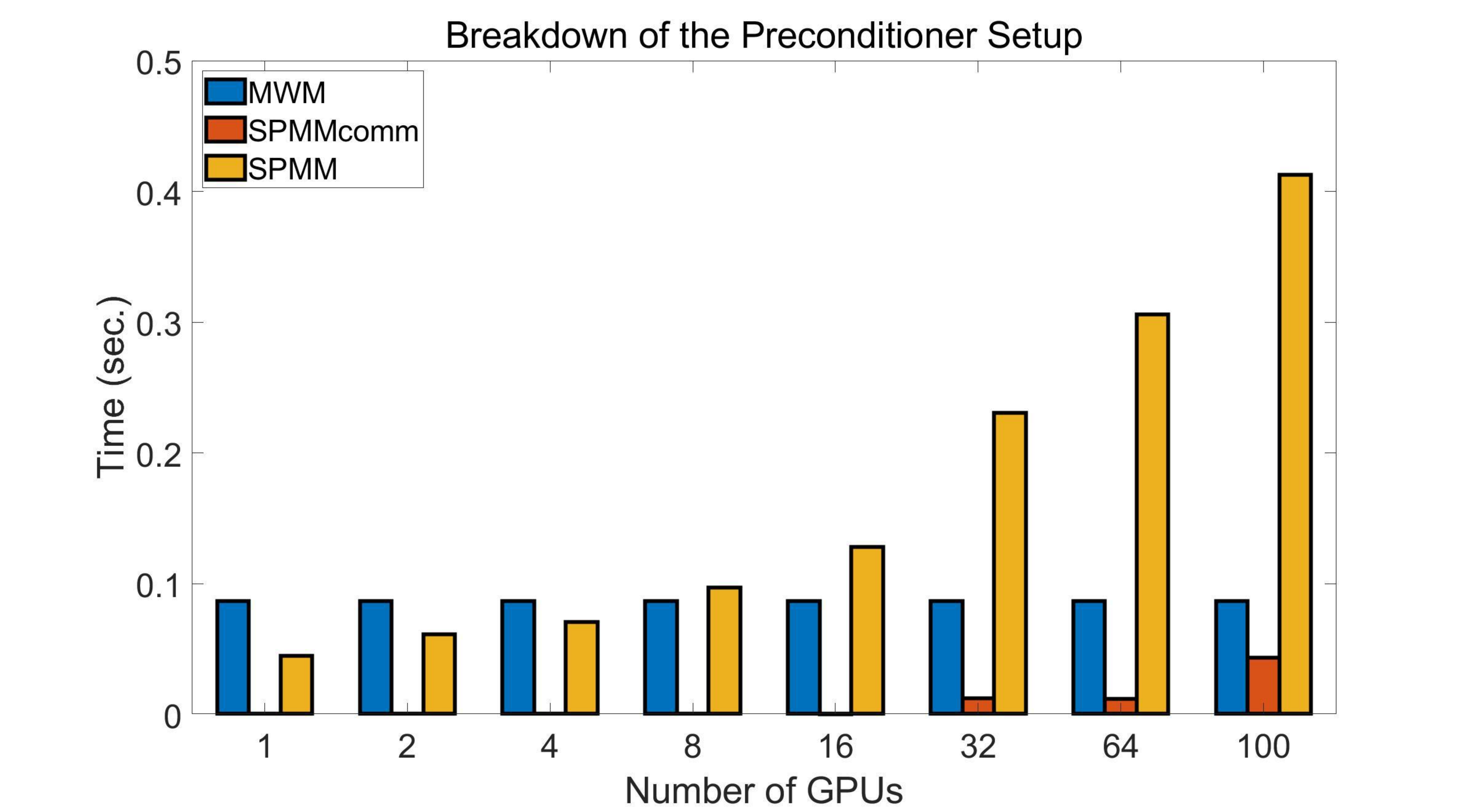}
\caption{Weak Scalability: Preconditioner Setup Time and Breakdown of Setup Time\label{fig-weak3}}
\end{center}
\end{figure*}

In Fig. \ref{fig-weak1}, we show operator complexity and number of iterations of {\em BCMG} versus {\em AMGX}. We can see that in all cases, $OPC$ is smaller for {\em BCMG} than {\em AMGX} and it is about $1.14$ for all the number of nodes. This confirm that our AMG hierarchy is generally cheaper than that of {\em AMGX} both in terms of memory footprint and in terms of the V-cycle application. On the other hand, in spite of the larger complexity, the number of iterations required by {\em AMGX} in the solve phase is always higher than that of {\em BCMG}. After an initial increase for both the solvers, they have a similar more stable behaviour, but the increase in the number of iterations for {\em AMGX} is $\sim 50 \%$ going from $1$ to $100$ nodes, whereas the increase for {\em BCMG} is of about $36 \%$, showing that {\em BCMG} produces hierarchies with better algorithmic scalability. This better quality and lower complexity of our preconditioner is confirmed by the solve time (see Fig. \ref{fig-weak2}). {\em BCMG} solve times are always significantly smaller than that of {\em AMGX}. In many cases {\em AMGX} requires a solve time that is double than that of {\em BCMG}. The more stable behaviour of the {\em BCMG} solve times for increasing number of resources and problem size is the result of the better scalability properties of our solver. Finally, if we look at the time per iteration, as expected from the values of the operator complexity, we see that {\em BCMG} always has smaller time than {\em AMGX}. On the other hand {\em BCMG} also shows a smaller increase ratio for increasing number of nodes, showing that all the computational kernels in the application phase of the preconditioner are efficiently implemented.
We observe from Fig.~\ref{fig-weak3} that, as expected, {\em AMGX} has a smaller setup time than \emph{BCMG} due to the very different implemented algorithms. On the other hand, our larger setup times are very well balanced by the significant smaller solve times.
Furthemore, we point out that in sparse linear solvers, a reasonably greater preconditioner setup time is tolerated if good convergence rates and parallel efficiency are obtained in the solve phase. Indeed, in time-dependent and/or non-linear applications, it is usually necessary to solve multiple linear systems with the same coefficient matrix or sequences of linear systems with slowly varying matrices that allow reusing of the preconditioner, therefore greater setup times are dampened over multiple solution steps.

For a deeper analysis of the performance of our setup phase,
a breakdown of the time spent in the two main computational kernels, i.e. {\em MWM} and {\em SpMM}, is reported in Fig.~\ref{fig-weak3}. The time of the measured overhead due to data communication required by the {\em SpMM} parallel computation is also shown (see SpMMComm bar in Fig.~\ref{fig-weak3}). We note that, as described in section~\ref{subsec:smsmprod}, this data communication is in charge of the CPU, while MWM computation runs on the GPU to make possible an overlap between MPI data communication and computation. We see that the extra-time needed for data communication, represented by the SpMMComm bar, is measured for a number of GPUs larger than 16 and, as expected, it increases for increasing number of GPUs.
It is worth to note that, when the problem size and the number of parallel nodes increase, as we already observed, there is an increase in the number of levels of the AMG hierarchy, then the amount of {\em MWM} and {\em SpMM} computations increase as well, while the dimension of the coarser matrices decreases. This is the reason of the increase in the execution time of the {\em SpMM} kernel and the related communication time.
%

\section{Conclusions}
\label{sec:concl}

In this paper, we discussed design choices and results of a hybrid MPI-CUDA implementation of a software library for solving large and sparse linear systems on multiple GPUs. The library design has been driven both at the algorithmic and at the implementation level by the need to exploit parallelism in each computational kernel of the linear solver. Best practices for hiding communication by overlapping data communication and computation have been applied, wherever possible. Weak scalability results and comparisons with a state-of-the-art software showed that our package is very promising for solving very large systems on heterogeneous parallel computers with multiple GPUs. Future work includes the extension of the library with new and robust highly parallel smoothers and coarsest solvers. The source code of the library is available at {\em https://github.com/bootcmatch/BootCMatchGX}.

  \section{Acknowledgments}

This work has been supported in part by the EC under the EuroHPC Project: {\em Towards EXtreme scale Technologies and Accelerators for euROhpc hw/Sw Supercomputing Applications for exascale (TEXTAROSSA)}, Horizon 2020 Program for Research and Innovation, ID: 956831  and by INdAM under GNCS Project: {\em 
InSANE: Investigating Sparse Algorithms in the post von Neumann Era
} CUP: E55F22000270001.

We gratefully acknowledge PRACE for granting us access to Piz Daint at ETH Zurich/CSCS, Switzerland.

The authors also thank Dario Pasquini, Post-Doc fellow at IAC-CNR, for helping in implementing some of the GPU kernels used in this work.

%
\vfill

\newpage
\section{Appendix}
\appendix

\section{How to use the solver}

Hereafter we briefly explain how a user can employ our linear solver within its application. Algorithm~\ref{mainTemplate} shows the pseudo-code of an example of usage. We remind the interested reader that the source code of the library is available at {\em https://github.com/bootcmatch/BootCMatchGX}. The repository contains all information needed to compile and run our solver. A sample program is included in the repository, and commands line parameters are described as well.

\begin{algorithm}
\caption{Example of Usage}
\label{mainTemplate}
\begin{algorithmic}[1]
\Statex $\textit{Prepare the input parameters $A$ and $b$}$
\Statex $\textit{The input matrix A is stored in CSR format}$
\State $\text{CSR} \; \texttt{*A} \gets \text{Generate or Read local matrix}$ \label{readMatrix}
\State $\text{vector}\langle vtype \rangle \; \texttt{*b} \gets \text{Generate or Read local rhs}$
\Statex $\textit{Read the configuration parameters}$
\State $\text{param} \; \texttt{p} \gets \text{Read Configuration File}$  \label{confParams}
\Statex $ \textit{Compute the solution} $
\State $\text{vector}\langle vtype \rangle \; \texttt{*x} \gets bcmgx(A, b, p, precflag)$ \label{bcmgx}
\end{algorithmic}
\end{algorithm}

\begin{algorithm}
\caption{Preconditioned Flexible Conjugate Gradient} \label{PCG-2}
\begin{algorithmic}[1]
\State Given $u_0$ and set $r_0 = b - Au_0$
\State $w_0 = d_0 = B r_0$ \label{line:prec0-2}
\State $v_0 = q_0 = A w_0$
\State $\alpha_0 = w_0^T r_0$
\State $\beta_0 = \rho_0 = w_0^T v_0$
\Statex
\State $u_{1} = u_0 + \alpha_0 / \rho_0 d_0$
\State $r_{1} = r_0 - \alpha_0 / \rho_0 q_0$
\Statex
\For{$i = 1,\dots$}
	\State $w_{i} = B r_{i}$ \label{line:precB-2}
	\State $v_{i} = A w_{i}$ \label{line:spmv-2}
	\Statex
	\State $\alpha_i = w_{i}^T r_{i}$ \label{line:stdot-2}
	\State $\beta_i = w_{i}^T v_{i}$
	\State $\gamma_i = w_{i}^T q_{i-1}$ \label{line:enddot-2}
	\Statex
	\State $\rho_i = \beta_i - \gamma_i^2 / \rho_{i-1}$
	\Statex
	\State $d_{i} = w_{i} - \gamma_i / \rho_{i-1} d_{i-1}$ \label{line:staxpy-2}
	\State $u_{i+1} = u_i + \alpha_i / \rho_i d_{i}$
	\Statex
	\State $q_{i} = v_{i} - \gamma_i / \rho_{i-1} q_{i-1}$
	\State $r_{i+1} = r_i - \alpha_i / \rho_i q_{i}$ \label{line:endaxpy-2}
\EndFor
\end{algorithmic}
\end{algorithm}

The first step for the usage of the solver is to prepare the input parameters, $A$ and $b$. The library provides a few functions to read and distribute, or generate on each GPU, the input matrix $A$ (line~\ref{readMatrix}). Then, solver parameters are read from the configuration file (line~\ref{confParams}), whose location can be specified through the command line. The library repository contains a sample of the configuration file with a short description of each parameter. 
Finally, the function \textit{bcmgx} (line~\ref{bcmgx}) sets up the preconditioner, if required, and compute the solution applying Algorithm~\ref{PCG-2}. 
The unpreconditioned version of the CG method corresponds to Algorithm~\ref{PCG-2} with $B=I$, where $I$ is the identity matrix, therefore it is obtained by a straightforward modification to the code which implements Algorithm~\ref{PCG-2}. 
In the sample program, the unpreconditioned CG can be set through the command line parameter \texttt{-p} and corresponds to the value $0$ for the {\em precflag} input parameter of {\em bcmgx}.
The configuration file used for the experiments discussed in this work is shown in Listing~\ref{amgsettings}.

\begin{lstlisting}[captionpos=b, abovecaptionskip=5pt, caption={Configuration File}, label={amgsettings}, frame=tb, basicstyle=\ttfamily\small, numbers=left, numberstyle=\small, numbersep=-8pt]
   NONE    % rhs file NONE if not present
   NONE    % sol file NONE if not present
   0       % bootstrap_type: 0 multiplicative, 
           % 1 symmetrized multi., 2 additive; 
           % NB: This is the composition rule 
           % when bootstrap is applied and more 
           % than 1 AMG hierarchy is setup
   1       % max_hrc, in bootstrap AMG, max 
           % hierarchies; NB: Here put 1 for 
           % single AMG component
   0.8     % desired convergence rate of the 
           % composite AMG; NB: This is not 
           % generally obtained if criterion 
           % on max_hrc is reached
   3       % matchtype: 3 Suitor
   2       % aggrsweeps; pairs aggregation 
           % steps. 0 pairs; 
           % 1 double pairs, etc ...
   0       % aggr_type; 0 unsmoothed, 
           % 1 smoothed (not yet supported)
   39      % max_levels; max number of levels 
           % built for the single hierarchy
   0       % cycle_type: 0-Vcycle, 1-Hcycle, 
           % 2-Wcycle
   4       % coarse_solver: 0 Jacobi, 1 FGS/BGS, 
           % 3 symmetrized GS, 4 l1-Jacobi
   4       % relax_type: 0 Jacobi, 1 FGS/BGS, 
           % 3 symmetrized GS, 4 l1-Jacobi
   20      % relaxnumber_coarse
   4       % prerelax_sweeps
   4       % postrelax_sweeps
   1000    % itnlim
   1.e-6   % rtol
   2000000 % mem_alloc_size
\end{lstlisting}

\section{Comparisons with related software}

\begin{figure*}[!ht]
\begin{center}
\includegraphics[width=0.48\textwidth]{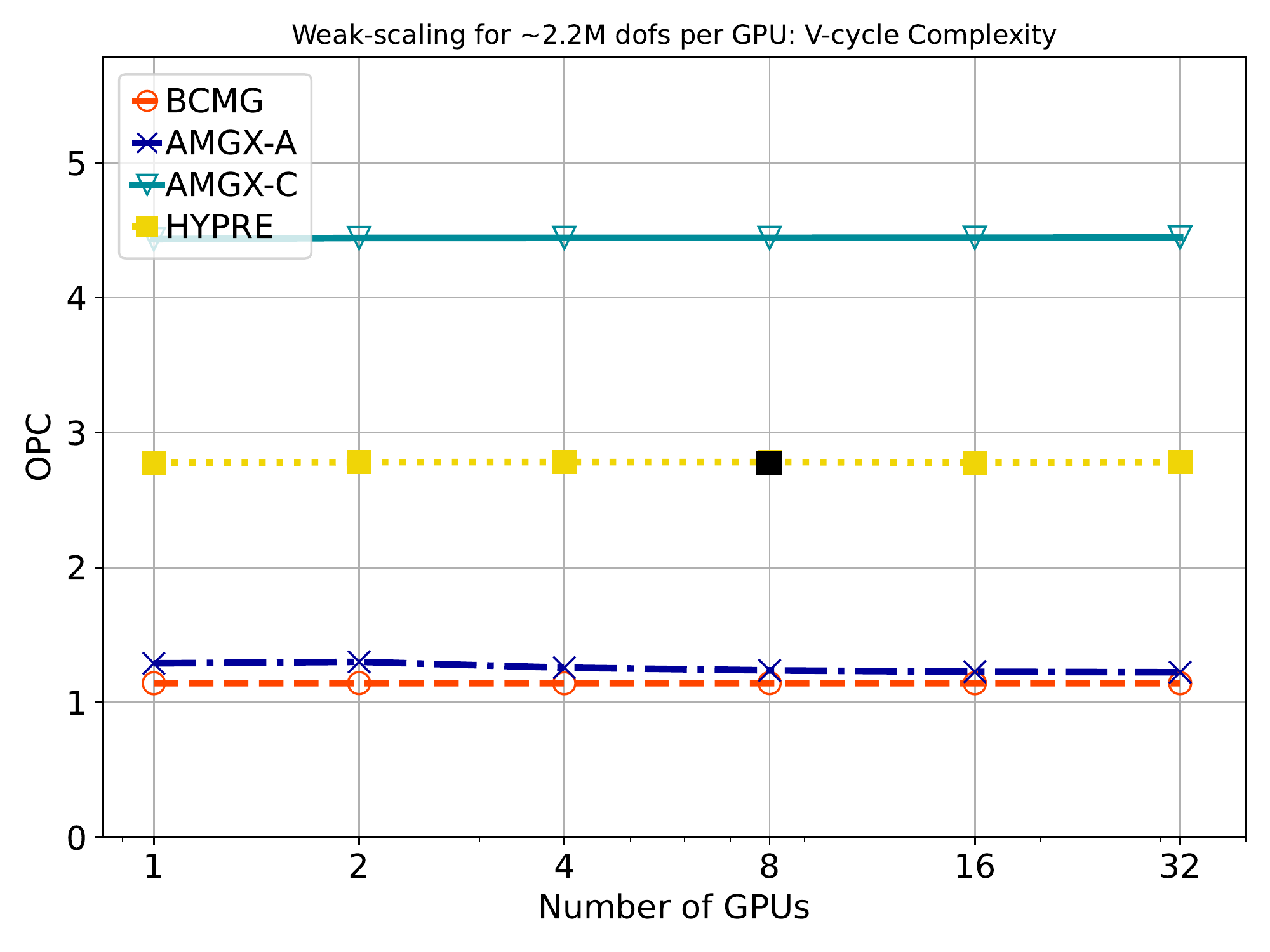}
\includegraphics[width=0.48\textwidth]{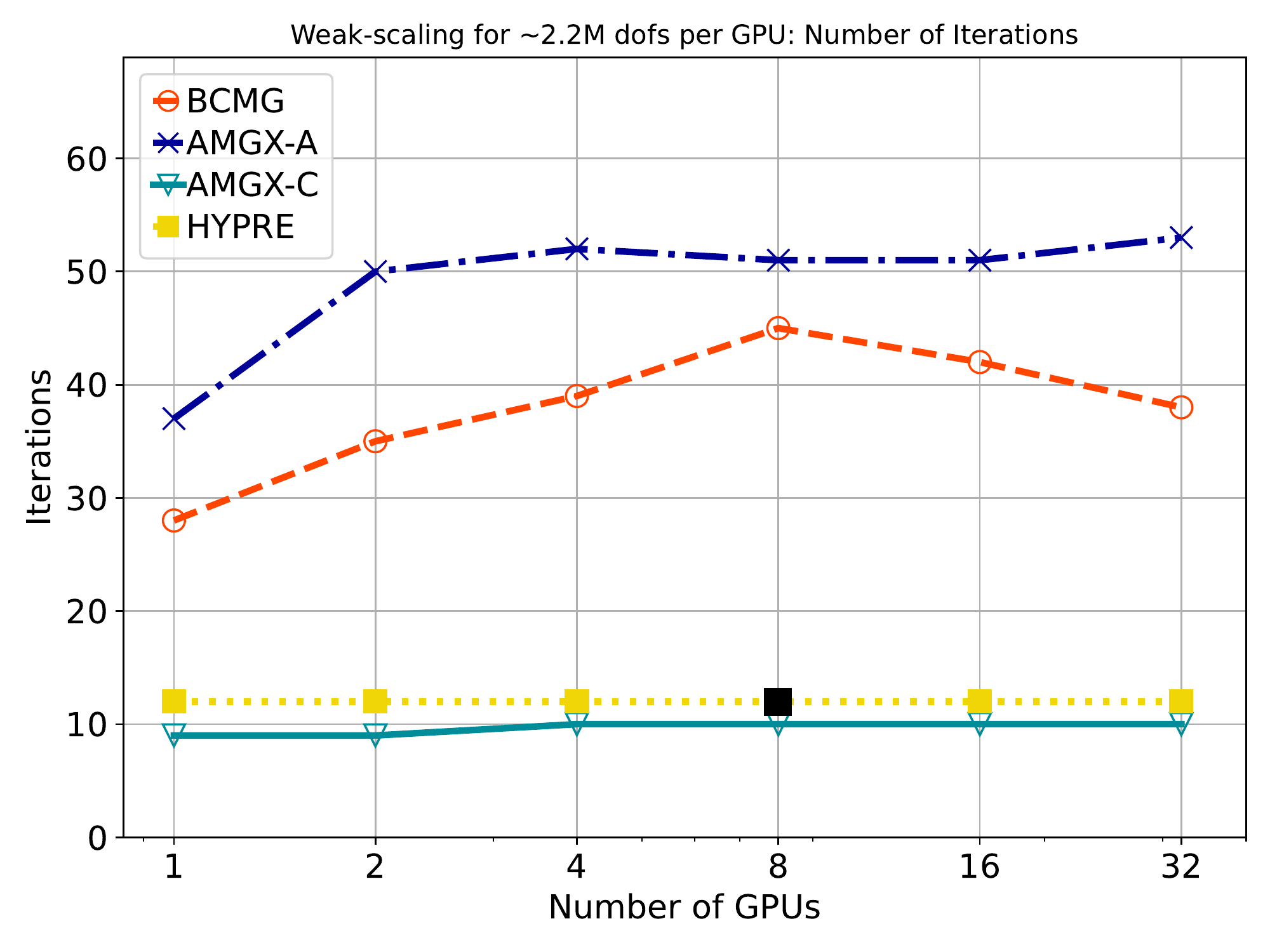}
\caption{Weak Scalability: Operator complexity (up) and Number of PCG Iterations (down)\label{fig-weak1-2}}
\end{center}
\end{figure*}

\begin{figure*}[!ht]
\begin{center}
\includegraphics[width=0.48\textwidth]{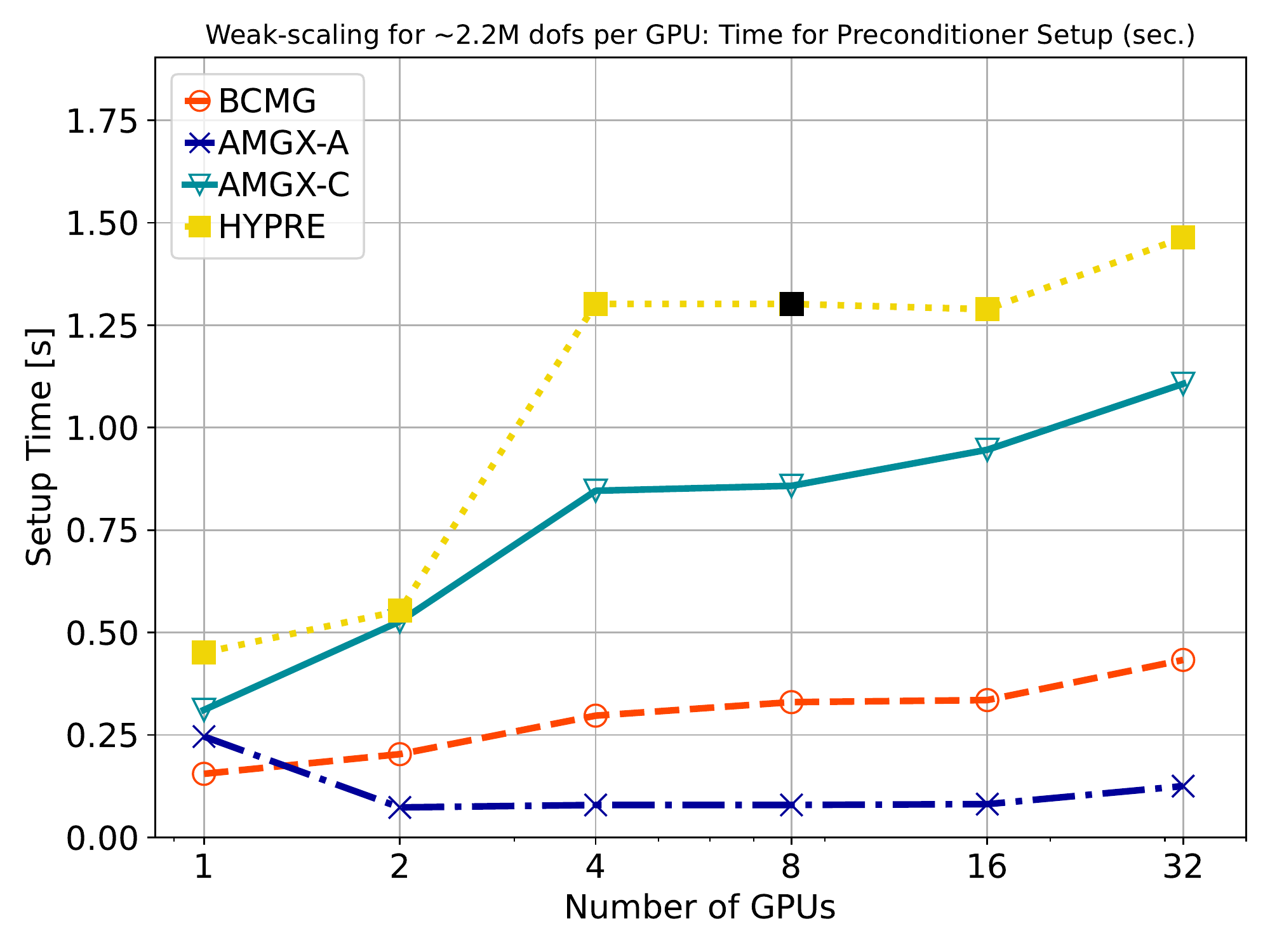}
\includegraphics[width=0.48\textwidth]{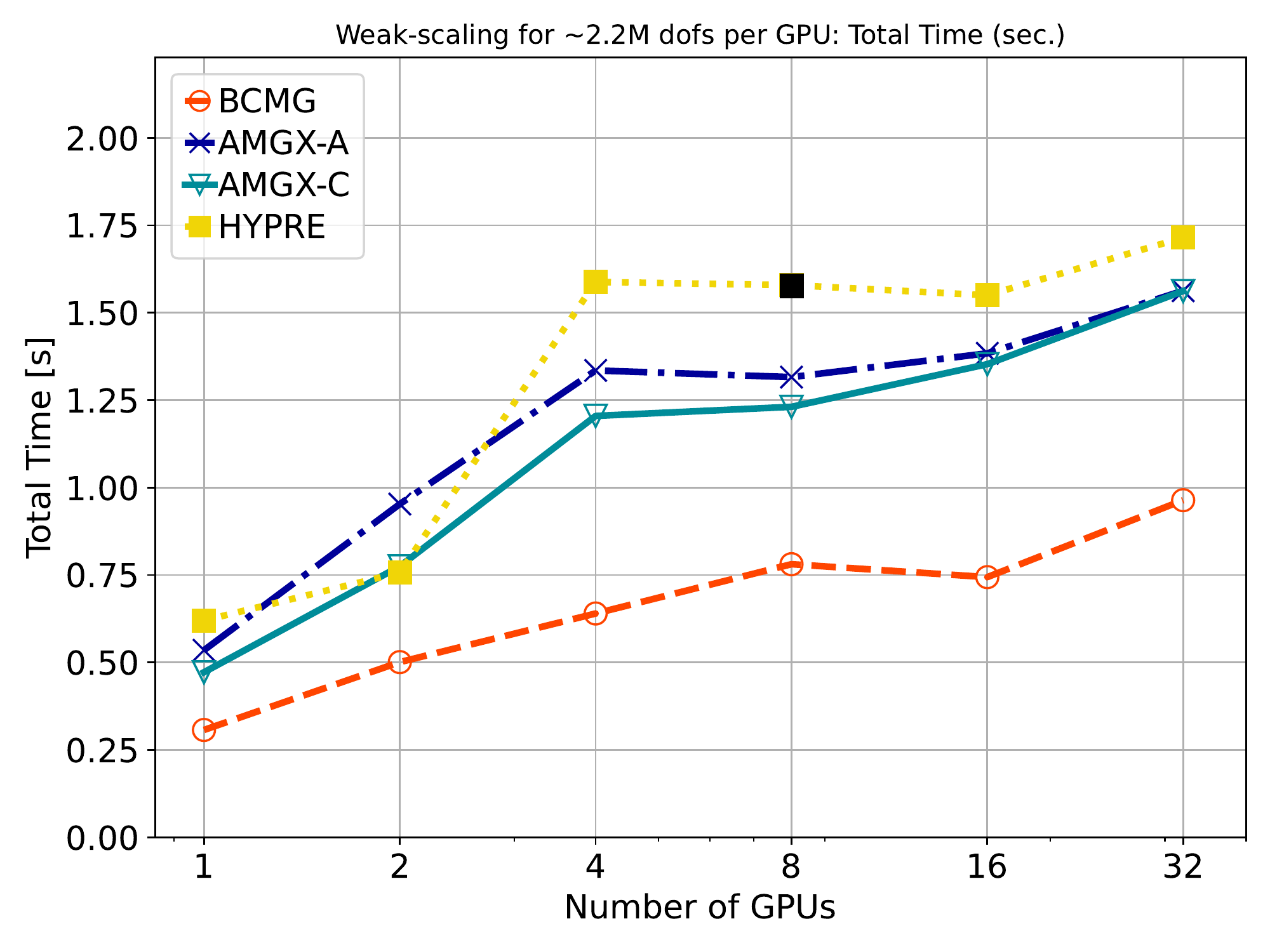}
\caption{Weak Scalability: Preconditioner Setup Time (up) and Total Time (down) \label{fig-weak3-3}}
\end{center}
\end{figure*}

In the following, we show comparison results obtained on the GPUs of the Marconi-100 supercomputer.
All the codes were compiled with the \texttt{gnu/8.4.0} suite, \texttt{CUDA 11.0} for GPU kernels, and linked against the \texttt{spectrum$\_$mpi/10.4.0} and \texttt{openblas/0.3.9} libraries.
We compared our {\em BCMG} solver both with the aggregation-based method of AmgX, already used in the paper (section 5.1 and section 5.2), hereby referred as {\em AMGX-A}, and with the AmgX implementation of the classical AMG method, explained in section 2.5 of the {\em AMGX Reference Manual} available at {\em https://github.com/NVIDIA/AMGX/blob/main/doc/}, hereby referred as {\em AMGX-C}. 
We used the NVIDIA AmgX library (rel. 2.2.0) already installed on Marconi 100.
Furthermore, we also installed on Marconi 100 the hypre library (rel. 2.26.0) available at {\em https://github.com/hypre-space/hypre}. For a fair comparison, among all the available AMG preconditioners, we selected the classical AMG method based on 
Parallel Maximal Independent Set (PMIS), which is the only coarsening method whose complete setup is currently supported on GPUs. The coarsening was applied using library default configuration to set up the AMG hierarchy. In the following, we refer to the solver based on hypre as {\em HYPRE}.
For all our tests, the same configuration of the V-cycle described in the paper is applied at each iteration of the PCG method. Therefore, we applied $4$ pre/post-smoothing iterations of $\ell_1-$Jacobi at intermediate levels and $20$ iterations of $\ell_1-$Jacobi at the coarsest level. Furthermore, we stopped PCG iterations when the relative residual in the $\ell_2$ norm is less than $10^{-6}$ or the number of iterations reaches the maximum value fixed to $1000$. Due to our grant restrictions, we were able to use up to $32$ GPUs of Marconi 100 and analyzed weak scalability of the solvers; the fixed matrix size per GPU is equal to $130^3 \approx 2.2 \times 10^6$. 
We point out that {\em HYPRE} does not run when $8$ GPUs are used and exits with an MPI error. A black square is put in the figures to emphasize that the line is the result of a constant interpolation from the previous value.
In Fig. \ref{fig-weak1-2}, we show operator complexity and number of iterations of the different solvers. We can see that in all cases, $OPC$ is smaller for {\em BCMG}, confirming that our AMG hierarchy is generally cheaper than that of the other preconditioners both in terms of memory footprint and in terms of floating-point operations needed for the V-cycle application. Similar operator complexity, as already observed in the paper, is obtained by {\em AMGX-A}, whereas very large operator complexity characterizes both {\em AMGX-C} and {\em HYPRE} due to their denser transfer operators set up in the AMG coarsening. The large operator complexity of {\em AMGX-C} and {\em HYPRE} leads, on one hand, to very good algorithmic scalability, as demonstrated by the small and very stable number of iterations for an increasing number of GPUs, and, on the other hand, to larger preconditioner setup time, as shown in Fig. \ref{fig-weak3-3} (up). The gain in the number of iterations allows the classical AMG methods of {\em AMGX-C} and {\em HYPRE} to have smaller solve time with respect to the aggregation-based AMG approaches albeit at the cost of a very large memory footprint. However, we observe that {\em BCMG} has solved time comparable to {\em AMGX-C} despite its smaller complexity. This demonstrates the effectiveness of our preconditioner and the overall implementation procedure. Finally, from  Fig. \ref{fig-weak3-3} (down), we can see that {\em BCMG} always obtains the best total time.
\begin{figure*}[!ht]
\begin{center}
\includegraphics[width=0.48\textwidth]{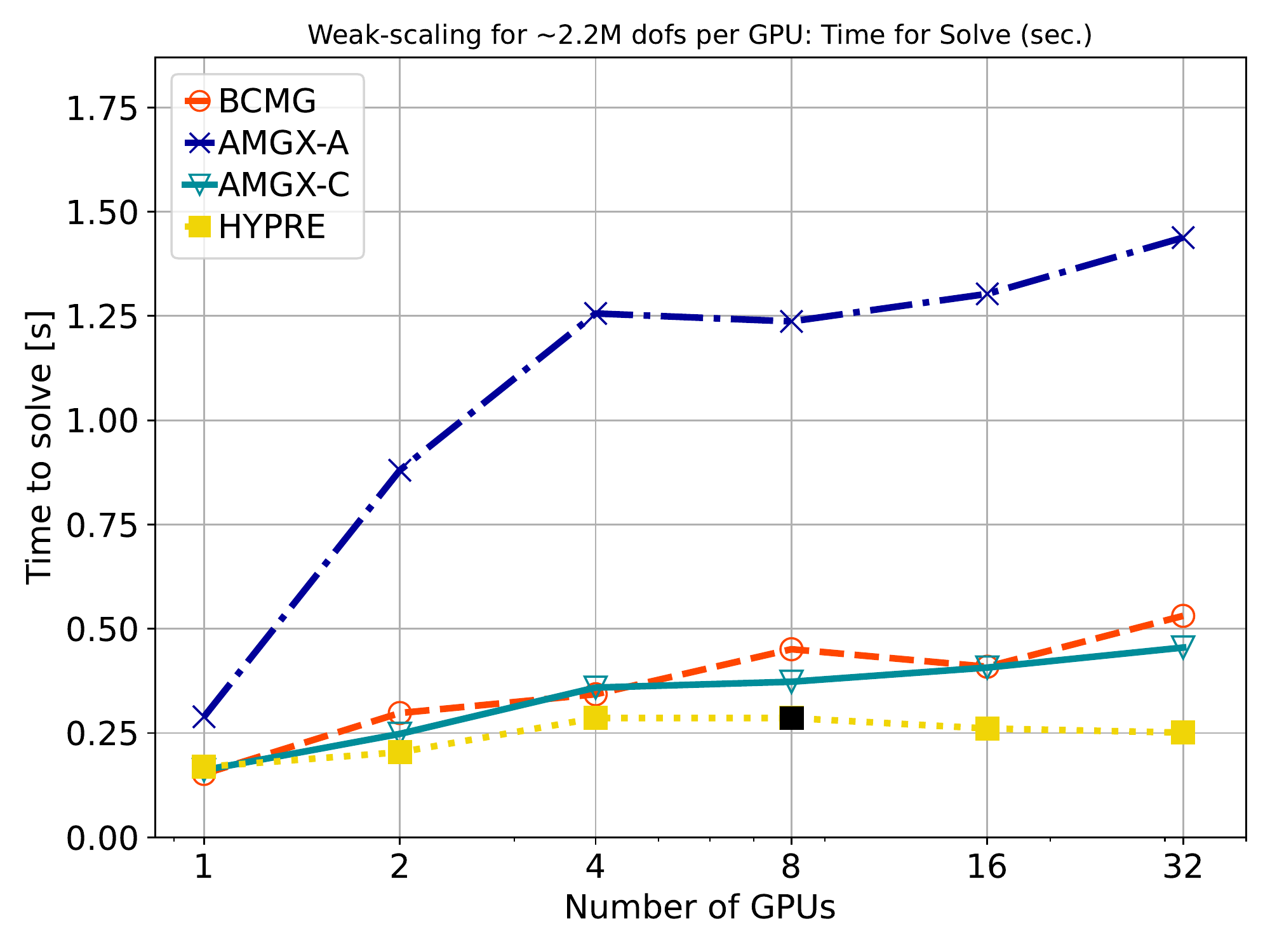}
\includegraphics[width=0.48\textwidth]{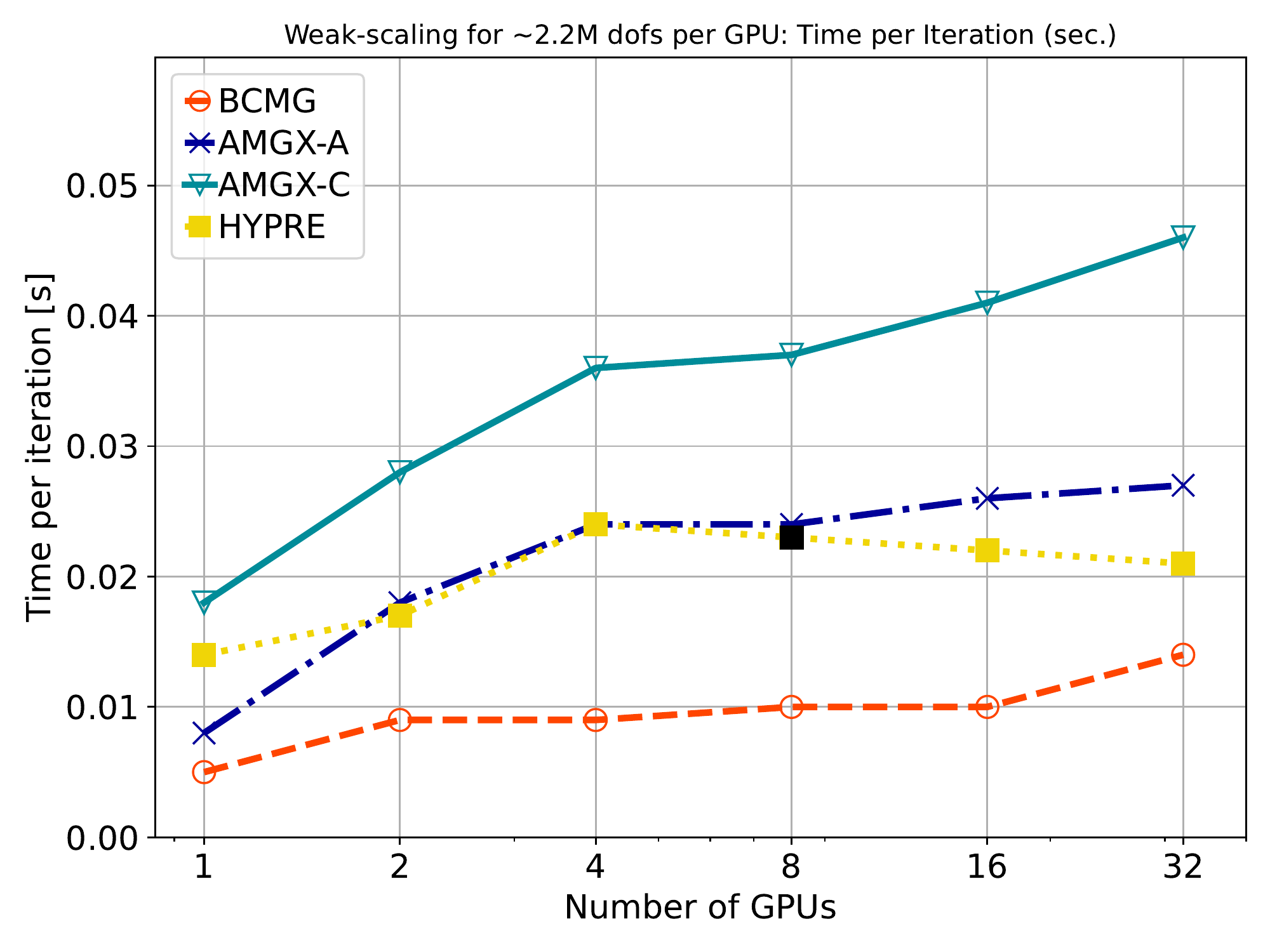}
\caption{Weak Scalability: Solve Time (up) and Solve Time per Iteration (down) \label{fig-weak2-2}}
\end{center}
\end{figure*}

\section{Example test and configuration files for AmgX Methods}

The experiments carried out in this work have been run by using the example program {\em amgx$\_$mpi$\_$poisson7} available in the examples folder of the AmgX distribution. This example program generates the data set described in the paper at section 5 and solve the related systems. The configuration file needed to run the {\em AMGXA} method and the {\em AMGXC} method are modifications of the {\em PCG$\_$AGGREGATION$\_$JACOBI.json} file~(Listing~\ref{pcg_aggregation}) and of the {\em PCG$\_$CLASSICAL$\_$V$\_$JACOBI.json} file~(Listing~\ref{pcg_classical}), respectively, which are distributed with the library. We report here both files for possible reproducibility needs.\\

\begin{lstlisting}[language=json, caption={Modified PCG$\_$AGGREGATION$\_$JACOBI.json}, label={pcg_aggregation}]
  {
    "config_version": 2,
    "determinism_flag": 1,
    "solver": {
        "preconditioner": {
            "print_grid_stats": 1,
            "algorithm": "AGGREGATION",
            "print_vis_data": 0,
            "selector": "SIZE_8",
            "solver": "AMG",
            "smoother": {
                "scope": "jacobi",
                "solver": "JACOBI_L1",
                "monitor_residual": 0,
                "print_solve_stats": 0,
                "relaxation_factor" : 1
            },
            "print_solve_stats": 0,
            "presweeps": 4,
            "coarse_solver": "JACOBI_L1",
            "coarsest_sweeps" : 20,
            "max_iters": 1,
            "monitor_residual": 0,
            "store_res_history": 0,
            "scope": "amg",
            "postsweeps": 4,
            "cycle": "V"
        },
        "solver": "PCG",
        "print_solve_stats": 1,
        "obtain_timings": 1,
        "max_iters": 1000,
        "monitor_residual": 1,
        "convergence": "RELATIVE_INI",
        "scope": "main",
        "tolerance": 1e-06,
        "norm": "L2"
    }
  }
\end{lstlisting}

\begin{lstlisting}[language=json, caption={Modified PCG$\_$CLASSICAL$\_$V$\_$JACOBI.json}, label={pcg_classical}]
  {
    "config_version": 2,
    "solver": {
        "preconditioner": {
            "print_grid_stats": 1,
            "print_vis_data": 0,
            "solver": "AMG",
            "smoother": {
                "scope": "jacobi",
                "solver": "JACOBI_L1",
                "monitor_residual": 0,
                "print_solve_stats": 0
            },
            "print_solve_stats": 0,
            "presweeps": 4,
            "interpolator": "D2",
            "coarse_solver": "JACOBI_L1",
            "coarsest_sweeps" : 20,
            "max_iters": 1,
            "monitor_residual": 0,
            "store_res_history": 0,
            "scope": "amg",
            "cycle": "V",
            "postsweeps": 4
        },
        "solver": "PCG",
        "print_solve_stats": 1,
        "obtain_timings": 1,
        "max_iters": 1000,
        "monitor_residual": 1,
        "convergence": "ABSOLUTE",
        "scope": "main",
        "tolerance": 1e-06,
        "norm": "L2"
    }
  }
\end{lstlisting}

\section{Configuration line for hypre}
Experiments with hypre have been carried out by using the $ij$ test program distributed with the library. In this case, we did not need any configuration file and the following options have been used for running: {\em -solver 1 -rlx$\_$down 18 -rlx$\_$up 18 -ns 4 -pmis}.

\end{document}